%% file: arxiv.tex
\documentclass[table,conference,review]{IEEEtran}
\IEEEoverridecommandlockouts
\usepackage{cite}
\usepackage{amsmath,amssymb,amsfonts}
\DeclareMathOperator*{\argmax}{arg\,max}
\usepackage{graphicx}
\usepackage{textcomp}
\usepackage{xcolor}

\usepackage{algorithm}
\usepackage[noend]{algpseudocode}
\usepackage{textcomp}

\usepackage{subcaption}
\usepackage{url}

\algblock{ParFor}{EndParFor}
\algnewcommand\algorithmicparfor{\textbf{parfor}}
\algnewcommand\algorithmicpardo{\textbf{do}}
\algnewcommand\algorithmicendparfor{\textbf{end\ parfor}}
\algrenewtext{ParFor}[1]{\algorithmicparfor\ #1\ \algorithmicpardo}
\algrenewtext{EndParFor}{\algorithmicendparfor}

\algnewcommand{\IIf}[1]{\State\algorithmicif\ #1\ \algorithmicthen}
\algnewcommand{\EndIIf}{\unskip\ \algorithmicend\ \algorithmicif}

\def\BibTeX{{\rm B\kern-.05em{\sc i\kern-.025em b}\kern-.08em
    T\kern-.1667em\lower.7ex\hbox{E}\kern-.125emX}}
\begin{document}

\title{Less is More: Faster Maximum Clique Search by Work-Avoidance}

\author{\IEEEauthorblockN{Hans Vandierendonck}
  \thanks{For the purpose of open access, the author has applied a Creative Commons Attribution 4.0 International (CC-BY-4.0) license to any Author Accepted Manuscript version arising.}
\IEEEauthorblockA{\textit{School of Electronics, Electrical Engineering and Computer Science} \\
\textit{Queen's University Belfast}\\
Belfast, United Kingdom \\
0000-0001-5868-9259}
}

\maketitle

\begin{abstract}
  The maximum clique (MC) problem is a challenging graph mining problem
  which, due to its NP-hard nature, can take a substantial amount
  of execution time.
  The MC problem is dominated by set intersection operations
  similar to Maximal Clique Enumeration, however it differs in requiring
  to find only a clique of maximum size.
  As such, key to the problem is to demonstrate efficiently
  that a particular part of the search space does not contain a maximum
  clique, allowing to skip over major parts of the search space.
  We present a number of techniques to optimize MC search in light
  of leaving major parts of the search space unvisited, including
  (i)~an efficient, lazily constructed graph representation;
  (ii)~filtering prior to initiating a detailed search;
  (iii)~efficient early-exit intersection algorithms;
  (iv)~exploiting algorithmic choice.
  These techniques result in a speedup of up to 38.9$\times$ compared
  to PMC, which is the most comparable algorithm, and a speedup
  up to 11$\times$ over MC-BRB.
\end{abstract}


\input{intro}
\input{background}
\input{method}
\input{eval}
\input{rela}
\input{concl}

\section*{Acknowledgment}
This work was supported in part by the
Engineering and Physical Sciences Research Council
(grant numbers EP/X029174/1, EP/X01974X/1 and EP/Z531054/1)
and the Department for the Economy Northern Ireland (grant number USI-226).

%\section*{References}
%\bibliographystyle{IEEEtran}
%\bibliography{../references.bib}


\end{document}

%% file: intro.tex
\section{Introduction}
\label{sec:intro}
The field of graph mining contains many challenging algorithms that are
of critical importance to data scientists.
In this work we focus on the Maximum Clique (MC) problem, which bears relevance
to many other graph mining problems.
A clique is a fully connected subgraph.
The MC problem is to find a clique of the maximum size in a graph.
The MC problem is NP-hard,
implying a worst-case exponential time complexity.
However, as with many NP-hard problems, many data sets can be processed
as efficiently as if using polynomial-time algorithms.

Exact algorithms are primarily based on the branch-and-bound pattern.
These algorithms maintain a set of candidate vertices and evaluate
what cliques can be built out of these candidates by recursively
applying two steps: (i)~selecting
a vertex (branching); and (ii)~assuring a large clique may be found (bounding).
The algorithms keep track of the incumbent clique, which
is the largest clique observed so far.
Algorithms vary by a number of strategies that affect performance.
The most important ones
are vertex ordering (often based on degeneracy~\cite{matula:83:degeneracy}),
filtering (ruling out candidate vertices),
pruning (ruling out entire sets of candidate vertices),
and inexact heuristic search to prime the incumbent clique.

Knowledge of the incumbent clique narrows the scope of the search.
Indeed, we are only interested to search for cliques that are larger still.
Based on this observation, a range of inferences
can be made on whether specific vertices and specific subgraphs
should be evaluated at all.
This leads to a number of effects that are common across MC algorithms:
(i)~as an incumbent clique of a large
size is known sooner, more vertices can be ruled out from the search
and the search will complete faster;
(ii)~as the incumbent clique grows larger, fewer neighbors of any vertex
remain relevant.
One may think of the incumbent clique as defining a subgraph of interest
that remains to be searched, which shrinks as the search progresses.

In this work we evaluate the implications of the
zone of interest on the data structures and algorithms that underpin
the MC problem.
The aim is to identify techniques that can be applied broadly
across MC algorithms.
We have identified 4 aspects where sizeable gains are possible.

\textbf{Graph Data Structure}:
Constructing a graph data structure takes time,
due to the size of the graphs we are evaluating,
due to relabelling of vertices in correspondence
to the vertex order defined by the MC algorithm,
and due to the use of data structures that support efficient intersections,
such as hash sets~\cite{blanusa:20:mce,besta:21:gms,vandierendonck:24:mce}.
A zone of interest that shrinks over the course of the computation
implies that a representation of the neighborhood is not needed for all
vertices, and where we do need them many of the neighbors will not
be of interest. We will motivate a
\textbf{lazy filtered hashed relabelled graph}
representation to take benefit of hashing and relabelling while
minimizing their overhead.

\textbf{Filtering}: 
Prior to engaging in the search of a candidate set, we propose
to filter out any vertices not in the zone of interest. 
Typically this is performed throughout the recursive search.
Advance filtering allows to reduce the size of the candidate set,
which has performance benefit on all subsequent set intersection operations,
filtering steps such as graph coloring, and others.
Moreover, it is possible in many cases to decide a candidate set should not
be searched at all.

\textbf{Algorithmic choice}: An immediate implication of filtering ahead
of the search is that low-degree vertices are removed.
Consequently, the remaining vertices are more tightly connected
and thus an induced subgraph with higher density is left.
Searching high-density subgraphs requires different algorithms than those
for searching low-density subgraphs,
which is an effect not widely capitalized on. We propose to use algorithmic
choice to accelerate search.

\textbf{Efficient set intersections}:
As with many graph mining problems~\cite{besta:21:sisa},
set intersections take up an important fraction of the execution time.
These set intersections are often conditional on finding a result that is
large enough given the context of the call.
An efficient algorithm for such intersections
has been introduced for the
Maximal Clique Enumeration problem~\cite{vandierendonck:24:mce}.
This algorithm exits early when it is clear that the result will not meet
the size target.
The MC problem requires different operations and we
introduce two new early exit intersection algorithms.

An extensive experimental evaluation on 28 graphs shows that
the proposed techniques
result in a median speedup of 3.21$\times$ and up to 38.9$\times$
over PMC~\cite{rossi:14:pmc}, a parallel MC algorithm,
a median speedup of 5.08$\times$ over dOmega~\cite{walteros:20:easy}
which solves the MC problem using $k$-vertex cover,
and 2.35$\times$ median speedup over MC-BRB~\cite{chang:19:brb},
an algorithm that transforms the MC problem to a sequence of $k$-clique
finding problems.
Moreover, we can solve graphs that cannot be solved
by these algorithms in the alotted time.

The remainder of this paper is organized as follows:
Section~\ref{sec:bg} presents background on the MC problem.
Section~\ref{sec:motiv} motivates this work.
Section~\ref{sec:method} describes our contributions.
Section~\ref{sec:eval} presents an experimental evaluation
and
Section~\ref{sec:rela} discusses further related work.

%% file: background.tex
\section{Background}
\label{sec:bg}
We assume a simple graph $G=(V,E)$ is determined by a set of vertices
$V$ and a set of edges $E\subseteq V\times V$. Edges are not directed, i.e.,
if $(u,v)\in E$ then also $(v,u)\in E$ for $u,v\in V$.
When $(u,v)\in E$ we say that $u$ is a neighbor of $v$,
or $u$ and $v$ are neighbors. The set of neighbors of a vertex constitutes
its neighborhood in the graph: $N_G(v)=\{u\in V: (u,v)\in E\}$.
The degree of a vertex is the size of its neigborhood.
The right-neighborhood of a vertex is determined in light of a vertex order
and is given by $N_G^{+}(v)=\{u\in N_G(v): u>v\}$ assuming $u>v$ evaluates
to true is $u$ is to the right of $v$.
If $S\subseteq V$, then $G[S]=(S,E\cap (S\times S))$
is the subgraph of G induced by the set $S$.

A clique $K$ is a completely connected graph without self-edges, i.e.,
for any $u,v\in V: (u,v)\in E$ iff $u\ne v$.
A maximal clique $C=(V_C,E_C)$ is a subgraph of a graph $G=(V,E)$,
i.e., $V_C\subseteq V$ and $E_C\subseteq E$ that is not contained in
any other clique. The maximum clique of a graph is the
largest subgraph of $G$ that forms a clique, i.e., the largest among all
maximal cliques of $G$. Multiple maximum cliques may exists.
The graph $G'=(V,E')$ is the complement of a graph $G=(V,E)$ if any edge
in $E$ is not present in $E'$ and vice versa. Self-edges are not allowed
in either $E$ or $E'$.

The degeneracy of a graph (also known as $k$-core) is a measure for how
sparse a graph is.
The $k$-core of a graph is the set of vertices that remain after repeatedly
removing all vertices with degree less than $k$~\cite{matula:83:degeneracy}.
A vertex has coreness $k$ if it belongs to a $k$-core but not to a $(k+1)$-core.
The degeneracy of a graph is the largest coreness of any of its vertices.
Coreness is linked to cliques as a vertex of coreness $k$ can be a member of at
most a $(k+1)$-clique.
The degeneracy of a graph is thus an upper bound to the maximum clique size.
Let $\omega(G)$ be the maximum clique size for $G$ and $d(G)$ the degeneracy of $G$
then $\omega(G)\le d(G)+1$. The clique-core gap $g(G)=d(G)+1-\omega(G)$ indicates the
gap between the maximum clique size and degeneracy.
Graphs with $g(G)=0$ are generally easier to solve~\cite{walteros:20:easy}.

\subsection{Algorithms for MC}
They key algorithmic idea for branch-and-bound
MC algorithms~\cite{carraghan:90:mc,tomita:03:mcq,ostergard:02:cliquer}
is to enumerate relevant subsets of vertices 
and to remember the largest clique encountered.
The largest clique observed so far is the incumbent clique, denoted $C^*$.
For each vertex $v\in V$, an initial clique $C=\{v\}$ is defined
and a set $P$ of candidate clique vertices is initialised
as the right-neighborhood of the vertex: $P\gets N_G(v)$.
Vertices $u$ are selected one by one from $P$ for addition to $C$
and the candidate set is reduced to retain
only those vertices that are neighbors to all vertices selected so far
($P\gets P\cap N_G(u)$).
The search is terminated when $|C|+|P|\le |C^*|$.

Vertex ordering is a critical parameter that
determines both the order of evaluating right-neighborhoods
and their composition.
The peeling order of the degeneracy
computation~\cite{matula:83:degeneracy} is generally
used~\cite{carraghan:90:mc,lu:17:rmc,ostergard:02:cliquer}.
This order guarantees that every right-neighborhood is no larger
than the coreness of the corresponding vertex~\cite{eppstein:10:mce}.
As such, generally small subgraphs are searched
and set intersections are applied to
sets smaller than the coreness of vertices~\cite{vandierendonck:24:mce}.

A variety of pruning rules can be applied, beyond testing the
size of the candidate set against the incumbent clique~\cite{wu:14:review}.
Graph coloring of $G[P]$, the subgraph induced by the candidate set, can reveal
the maximum size of a clique contained in this subgraph, as a clique
of size $k$ requires $k$ colors~\cite{babel:90:mc,tomita:03:mcq}.
Search can be terminated when $|C|+\chi(G[P])\le |C^*|$ where
$\chi(G[P])$ is the chromatic number of $G[P]$.
SAT-based reasoning can identify infeasible candidate sets based by
deducing conflicting constraints on the connections in $G[P]$ and
the size of the incumbent clique~\cite{li:17:momc,sansegundo:23:clisat}.

Heuristic search, a greedy non-optimal search for large cliques,
has been shown to be effective to identify a large incumbent clique
fast~\cite{rossi:14:pmc}. This in turn accelerates filtering and pruning.
Heuristic search can be executed before
computing the degeneracy order~\cite{chang:19:brb}
or after~\cite{rossi:14:pmc}. The former can also reduce the time
for degeneracy computation as we can remove all low-degree vertices
from this computation.

Various authors have experimented with solving the MC problem
using different algorithms,
such as minimum vertex cover~\cite{walteros:20:easy},
maximum independent set~\cite{li:10:maxsat},
and $k$-clique search~\cite{chang:19:brb}.
These approaches work on the premise that these algorithms are more
efficient given how the subproblems have been defined.

\subsection{Equivalent Problems}
The MC problem is equivalent to other problems on graphs.
The minimum vertex cover (MVC) problem is to find the minimum set of vertices
such that at least one of the vertices incident to each edge are included
in the cover.
The minimum vertex cover of a graph is the complement of the maximum
clique of the complement of the graph.
The $k$-vertex cover ($k$-VC) problem decides if a vertex cover of size $k$
exists.
The maximum independent set (MIS) $S$
is the largest set $S\subseteq V$ such that no
two neighboring vertices are included, i.e., $(u,v)\not\in E$ if $u,v\in S$.
The maximum independent set of a graph is a clique of the complement graph.

\begin{figure}[!t]
  \subfloat[Clique-core gap zero]{\label{fig:mustmay:gapz}
    \centering
    \includegraphics[width=.45\columnwidth]{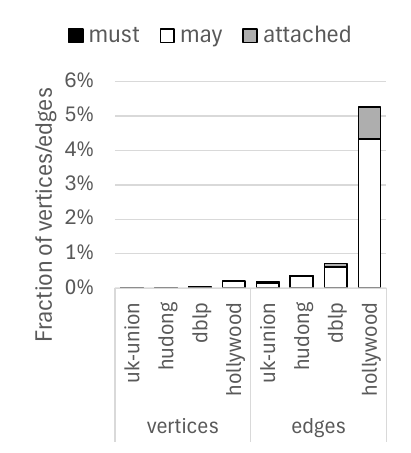}
  }%
  \subfloat[Clique-core gap non-zero]{\label{fig:mustmay:gapnz}
    \centering
    \includegraphics[width=.45\columnwidth]{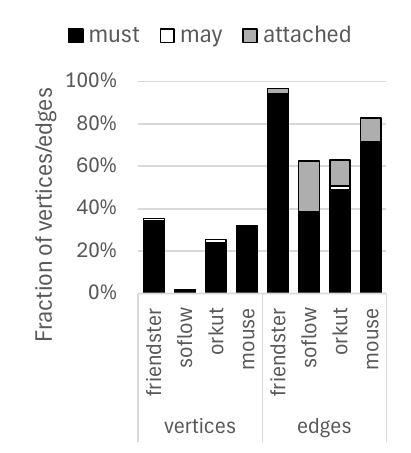}
  }%
  \caption{Characterization of \emph{must} and \emph{may} subgraphs.
    The \emph{must} subgraph is strictly contained in the \emph{may} subgraph,
  whose edges are a subset of the \emph{attached} edges.}
  \label{fig:mustmay}
\end{figure}

\section{Motivation}
\label{sec:motiv}
\label{sec:obs}
The presence of a zone of interest
that defines what parts of the graph are to be searched
creates optimization opportunities.

\subsection{Only A Small Fraction of the Graph Is Accessed}
We introduce the notion of \emph{may} and \emph{must} vertices
to describe the zone of interest at the end of the search.
We define as vertices that \emph{must} be inspected those vertices
whose coreness $c(v)>\omega(G)-1$ (a pre-condition to being contained in
a clique larger than $\omega(G)$),
and vertices that \emph{may} be inspected as the vertices whose coreness
$c(v)\ge\omega(G)-1$.
The \emph{may} and \emph{must} edges are those
edges in the corresponding induced subgraph.

Figure~\ref{fig:mustmay} shows the fraction of the vertices and edges
that may or must be inspected during MC search.
These are computed
after the maximum clique size is known, for the purpose of illustration.
Indeed, in the best of situations, we may immediately find the maximum clique,
which can succeed by investigating all \emph{may} vertices. In practice,
some lower-coreness vertices will be evaluated before the \emph{may} vertices.
When the maximum clique is found, we still need to rule out the existence of
a larger maximum clique, which requires evaluation of the \emph{must} vertices.

Graphs with a clique-core gap of zero ($\omega(G)=d(G)+1$)
(Figure~\ref{fig:mustmay:gapz})
such as \emph{uk-union} have an empty \emph{must} graph
because no vertex has a coreness larger than $\omega(G)-1$.
The search may be terminated after evaluating only a few vertices.
For graphs with a non-zero clique-core gap (Figure~\ref{fig:mustmay:gapnz})
the \emph{must} subgraph
is a more substantial subgraph of $G$.
This indicates a much greater computational
task even once the MC has been identified.
Nonetheless, less than 40\% of the vertices in these graphs and as
few as 40\% of the edges need to be searched.
As such,
\textbf{the graph representation only needs to provide
access to a small fraction of vertices and their neighbors}.
This motivates the construction of a graph representation that selectively
represents only vertices of interest and their neighbors.

Those vertices that we do search also attach to vertices
outside the \emph{may} subgraph
(Figure~\ref{fig:mustmay}, column ``attached'').
This could account for up to 20\% of the edges in the graph, and
more than a third of the neighbors of the \emph{may} vertices in the
case of the soflow graph.
As such,
\textbf{the majority of neighborhoods include
vertices that are already ruled out by knowledge of the incumbent clique}.
An efficient graph representation should limit the neighborhoods
to take into account
the incumbent clique size, which is evolving throughout the execution.

\subsection{Implication on Graph Representation}
As the incumbent clique increases in size throughout the computation,
the zone of interest shrinks until it becomes as small as the \emph{must}
subgraph. It is important to \textbf{adapt the graph representation}
along with the shrinking zone of interest.
In particular, as the neighborhood of a vertex is searched
later in the computation, a higher number of its neighbors can be ruled
out from the search. As such, we propose
\textbf{to construct the graph representation lazily, applying filtering
  at the time of construction}.

We will also \textbf{memoize representations that are
relatively expensive to construct}, such as relabelling of vertices to
reflect the vertex order applied by the MC algorithm, and to construct
data structures that allow efficient set intersection operations
such as hash sets.
The state of the art approaches this by making suboptimal choices:
either working from the
original representation and repeatedly relabelling vertices whenever necessary,
e.g.,~\cite{chang:19:brb}, %,others},
or creating a relabelled representation that may remain largely unused,
e.g.,~\cite{rossi:14:pmc,tomita:03:mcq,tomita:10:mcs,carraghan:90:mc}.
A lazy approach mitigates the up-front cost while avoiding repeated
transformations.

\subsection{Reduction to the Zone of Interest}
Before searching a particular candidate set, it is important
to determine whether it is in the zone of interest.
This is obvious in many cases, e.g., when a vertex has a degree or
coreness that is less than the size of the incumbent clique.
In other cases it is less easy to tell.
MC algorithms will discover whether a candidate set is in the zone of interest
during the search.
However, we propose to \textbf{filter candidate sets ahead of the recursive
  search} to reduce them to the zone of interest.
This has three benefits: (i)~the recursive search need not repeatedly
established the irrelevance of filtered-out vertices;
(ii)~in many cases we can rule out the need for any search at all;
(iii)~the candidate set is reduced in size by removing vertices that are
no longer of interest, implying that all operations including vertex selection,
set intersections, and filtering rules based on graph coloring operate
on smaller subgraphs.

\subsection{The Density of Searched Subgraphs is High}
Filtering increases the density of the subgraph.
Indeed, typically the lowest-degree vertices are excluded,
which increases the average degree and thus the density for the remaining
subgraph.
Subgraphs often have densities in excess of 90\% and
the \emph{must} and \emph{may}
subgraphs are up to six orders of magnitude more dense than the whole graph.
Solving very high-density subgraphs poses challenges as (i)~the number
of cliques to consider may be high, and (ii)~intersections are
applied to sets which are large relative to the size of the subgraph.
We propose to use \textbf{algorithmic choice} between solving the MC problem
on a dense subgraph, versus solving the $k$-VC problem on the complement
subgraph, which is sparse.

%% file: method.tex
\section{Method}
\label{sec:method}
\begin{algorithm}[t]
  \caption{High-level overview of the LazyMC algorithm.}
  \label{algo:lazy}
  \begin{algorithmic}[1]
    \small
    \Function{LazyMC}{$G$}
    \State $C^*=\{\}$ \algorithmiccomment{incumbent clique, global variable}
    \State $DegreeBasedHeuristicSearch(G)$
    \State $c=KCore(G,|C^*|)$\algorithmiccomment{c[v]: coreness of v if $d(v)\ge |C^*|$}
    \State $o=DetermineSortOrder(G,c)$
    \State $H=LazyGraph(G,o,c)$
    \State $CorenessBasedHeuristicSearch(H,c)$
    \State $SystematicSearch(H,c)$
    \EndFunction
  \end{algorithmic}
\end{algorithm}
We define several techniques to speedup the MC problem and demonstrate these
in a new MC algorithm called LazyMC.
The general structure of LazyMC follows established ideas for branch-and-bound
MC algorithms using commonly proven techniques.
Algorithm~\ref{algo:lazy} describes the algorithm at a top level,
combining two heuristic searches, vertex ordering based on degeneracy
and a systematic search. These steps are explained in subsequent sections.

% LFHRG
\subsection{A Lazy Filtered Hashed Relabelled Graph}
The majority of the MC search operates on the reordered graph, where the
right-neighborhoods of each vertex have been made small based on the
coreness order.
As such, the reordered neighborhoods of vertices is required by the
search. Based on the observations in Section~\ref{sec:obs}, we have designed
a \emph{lazy filtered hashed relabelled graph}
representation to minimize the overhead of constructing a relabelled graph
and querying the neighbor sets.
This representation achieves several goals:
\begin{itemize}
\item \textbf{Relabelling}: apply relabelling (which involves
  random memory accesses with a high probability of cache misses)
  only when needed and memoizing the result.
\item \textbf{Lazy construction}: postponing the construction of neighborhoods
  to the time they are needed, avoiding overheads for those that are never used.
\item \textbf{Filtering neighborhoods}: removing those neighbors from the
  representation that can be ruled out on the basis of the incumbent clique
  size at the time of construction. This accelerates future operations on the
  neighborhoods, in particular set intersections.
\item \textbf{Hashed sets}: constructing hashed sets which enable faster
  intersection operations.
\end{itemize}
Algorithm~\ref{algo:graph} gives the idea of the representation: the
interface to a lazy graph data structures allows to query for
a hash set representation, a sorted array representation or either of these.
These representations are constructed when queried.

Either or both a sorted array representation and a hash set representation
may be constructed for each vertex.
Their sizes may differ depending on the size of the incumbent
clique at the time that they are constructed. This does not affect correctness,
as any discrepancies between them relate to vertices that have no further
impact on the search.

Which representation is created depends on the context.
A sorted array representation is constructed when the elements of the set
will be iterated over one by one.
This happens when the neighborhood of a vertex is queried at the top level
of the search.
A hash representation will be constructed when the induced subgraph of
a neighborhood is cut out for an MC search.
Other contexts can work with either representation. If none exists,
then a hash representation is created for high-degree vertices
(degree over 16), and a sorted array is created otherwise.
If both exist, then intersections will be performed using the hash set
rather than the sorted array representation.

This graph representation is shared by all threads. 
Concurrency is managed with double-checked locking
as each neighborhood is read-only after construction.
This ensures fast access to already constructed sets
and maintains the simplicity of mutual exclusion while neighborhoods
are constructed.

\begin{algorithm}[t]
  \caption{Principles behind the lazy graph data structure.}
  \label{algo:graph}
  \begin{algorithmic}[1]
    \small
    \Function{LazyGraph}{$G$,$o$,$c$}
    \State $f\gets$ array of $|V|$ integers, initially zero
    \State $h\gets$ array of $|V|$ hash tables, initially empty
    \State $s\gets$ array of $|V|$ arrays, initially empty
    \State $l\gets$ array of $|V|$ locks
    \State \Return Construct LazyGraph object with $G$, $o$, $c$, $f$, $h$, $s$, $l$
    \EndFunction
    \Function{GetHashedNeighborhood}{LazyGraph $H$,$v$}
    \If{$H.f[v]\& 1==0$}\algorithmiccomment{No hash version present}
        \State \Return CreateHashedNeighborhood($H$,$v$)
    \Else
    \State \Return $H.h[v]$
    \EndIf
    \EndFunction
%    \Function{GetAnyNeighborhood}{LazyGraph $H$,$v$}
%    \If{$H.f[v]==0$}\algorithmiccomment{No versions present}
%        \State $v_o\gets H.o.remap\_to\_original(v)$
%        \If{$|N_{H.G}(v_o)|\ge 16$}
%        \State \Return CreateHashedNeighborhood($H$,$v$)
%        \Else
%        \State \Return CreateArrayNeighborhood($H$,$v$)
%        \EndIf
%    \ElsIf{$H.f[v]\&1==1$}\algorithmiccomment{Hashed version present}
%        \State etc
%        \EndIf
%    \EndFunction
    \Function{CreateHashedNeighborhood}{LazyGraph $H$,$v$}
    \State $H.l[v].lock()$
    \If{$H.f[v]\& 1==0$}\algorithmiccomment{Double-checked locking}
        \State{}\algorithmiccomment{Lazy initialization of neighbor hash set}
        \State $v_o\gets H.o.relabelled\_to\_original(v)$
        \State $H.h[v].initialize(|N_{H.G}(v_o)|)$\algorithmiccomment{Reserve memory}
        \For{$u_o\in N_{H.G}(v_o)$}\algorithmiccomment{Remap vertices from $G$ to $H$}
        \State $u\gets H.o.original\_to\_relabelled(u_o)$
        \If{$H.c[u]\ge |C^*|$}\algorithmiccomment{Lazy filtering by coreness}
        \State $H.h[v].insert(u)$
        \EndIf
        \EndFor
    \EndIf
    \State $H.f[v]\gets H.f[v] | 1$\algorithmiccomment{Set flag initialized; atomic update}
    \State $H.l[v].unlock()$
    \State \Return $H.h[v]$
    \EndFunction
  \end{algorithmic}
\end{algorithm}

\subsection{Early-Exit Set Intersections}
Both the degree-based heuristic search and filtering
require to identify the highest-degree
vertex in an induced subgraph: given a vertex subset $S\subseteq V$,
the degrees in the induced subgraph are given by $d_S(v)=|N_G(v)\cap S|$
for $v\in S$. The goal is to find $\argmax_{v\in S} d_S(v)$.
Suppose that $d_S(v_0)=k$ for some $v_0\in S$.
Then, when evaluating vertex $v_1\in S$, the outcome of this
computation depends only on the value $|d_S(v_1)|$
if $|d_S(v_1)|>k$. Otherwise, it is irrelevant as $v_1$ is not the
unique maximum-degree vertex in $S$.

Early-exit set intersections make use of this property and exit the intersection
operation early when it becomes clear that the result set cannot be
of the required size.
The operation \emph{intersect-size-gt-val}$(A,B,\theta)$ returns
$|A\cap B|$ if this is larger than $\theta$.
Otherwise, any smaller size or an error code is returned.
This can be effected by tracking how many elements
of $A$ are not in the intersection of $A\cap B$ and by exiting with
a failure code if
this number becomes larger than $|A|-\theta$~\cite{vandierendonck:24:mce}.

\begin{algorithm}[t]
  \caption{intersect-gt: intersects the sets
    $A$ and $B$ and places the result in size $C$ and returns its size.
    It may return -1 when the size is $\theta$ or less.
    $A$ is provided as array of length $n$
    and $B$ as a hash table with $m$ elements.}
    \label{algo:gt}
  \begin{algorithmic}[1]  
    \small
    \Function{intersect-gt}{$A$, $B$, $C$, $\theta$}
    \State $n=|A|$; $m=|B|$
    \If{$n<\theta$ \textbf{or} $m<\theta$} \Return -1
    \EndIf
    \State $h\gets n-\theta$\algorithmiccomment{number of missing elements tolerated}
    \For{$a \gets 0$ to $n-1$}
        \If{$!B.contains(A[a])$}
            \State $h\gets h-1$
            \If{$h\le 0$} \Return $-1$\algorithmiccomment{exit early}
            \EndIf
        \Else
            \State $C[a+h-(n-\theta)]\gets A[a]$
        \EndIf
    \EndFor
    \State \Return $h + \theta$
    \EndFunction
  \end{algorithmic}
\end{algorithm}

\begin{algorithm}[t]
  \caption{intersect-size-gt-bool: returns true if the
    size of the intersection of $A$ and $B$
    is larger than $\theta$ and false otherwise.
    $A$ is provided as array of length $n$
    and $B$ as a hash table with $m$ elements.}
    \label{algo:gt-bool}
  \begin{algorithmic}[1]  
    \small
    \Function{intersect-size-gt-bool}{$A$, $B$, $\theta$}
    \State $n=|A|$; $m=|B|$
    \If{$n\le\theta$ \textbf{or} $m\le\theta$} \Return false
    \EndIf
    \State $h\gets n-\theta$\algorithmiccomment{number of missing elements tolerated}
    \For{$a \gets 0$ to $n-1$}
        \If{$!B.contains(A[a])$}
            \State $h\gets h-1$
            \If{$h\le 0$} \Return $false$\algorithmiccomment{exit early}
            \EndIf
        \ElsIf{$h>n-a-1$} \Return $true$\algorithmiccomment{exit early}
        \EndIf
    \EndFor
    \State \Return $h > 0$
    \EndFunction
  \end{algorithmic}
\end{algorithm}

We define two related early-exit set intersection operations.
\emph{intersect-gt}$(A,B,C,\theta)$
returns the exact intersection result in the buffer $C$ when its
size is larger than $\theta$. If the result is smaller, any partial
result set or an error code may be returned. This operation is used
in heuristic search.

The algorithm is illustrated in Algorithm~\ref{algo:gt}.
The variable $h$ tracks the number of elements of $A$ that we may omit
and still have an intersection larger than $\theta$. Each time an element
of $A$ is not in $B$, we decrement this number. Otherwise, we place
the element in $C$.

The operation \emph{intersect-size-gt-bool}$(A,B,\theta)$
returns the boolean value $|A\cap B|>\theta$.
This is useful during filtering, where we need to determine if a vertex
has a sufficient number of neighbors in an induced subgraph such that
it can potentially be part of a clique larger than the incumbent clique.

When checking if the size of the intersection is larger than a threshold,
we can employ two early exits, corresponding to having obtained
sufficient evidence that the response is either true or false.
Algorithm~\ref{algo:gt-bool} exits early if so many elements
of $A$ are not appearing in $B$ that it will become impossible to find
$\theta+1$ elements in total. This is similar to \emph{intersect-gt}
and \emph{intersect-size-gt-val}.
There is a second early exit that checks if $h>n-a-1$, i.e., if
we can tolerate to omit more elements of $A$ (given by $h$)
than those remaining to check for membership in $B$ (given by $n-a-1$)
then clearly the intersection size will be larger than $\theta$.
Note that the $-1$ corrects for the element $A[a]$ also being included
in the intersection.
This second early exit allows \emph{intersect-size-gt-bool} to exit an
intersection for a very large set early, where \emph{intersect-size-gt-val}
needs to complete the operation to establish the precise intersection size.

\subsection{Heuristic Search}
\begin{algorithm}[t]
  \caption{Degree-based heuristic search.}
  \label{algo:dheur}
  \begin{algorithmic}[1]
    \small
    \Function{DegreeBasedHeuristicSearch}{$G$}
    \State Identify top-K vertices with highest degree
    \ParFor{$v\gets$ one of top-K}
        \State $N\gets \{u\in N_G(v): |N_G(u)|\ge |C^*|\}$
        \State $C=\{v\}$
        \Repeat
            \State Using \emph{intersect-size-gt-val}:
            \State $u\gets \argmax_{w\in N} |N\cap N_G(w)|$\algorithmiccomment{\emph{intersect-gt}, $\theta=$running maximum}
            \State $C\gets C\cup \{u\}$
            \State $N\gets N\cap N_G(u)$
        \Until{$|N|==0$}
        \If{$|C|>|C^*|$} \State $C^*\gets C$
        \EndIf
    \EndParFor
    \EndFunction
  \end{algorithmic}
\end{algorithm}%
\begin{algorithm}[t]
  \caption{Coreness-based heuristic search}
  \label{algo:cheur}
  \begin{algorithmic}[1]
    \small
    \Function{CorenessBasedHeuristicSearch}{$H$,$o$}
    \ParFor{$k\gets degeneracy+1$ downto 1}
        \State $v\gets $ lowest-numbered vertex with $c[v]=k$
        \State $N\gets N_G^{+}(v)$
        \State $C=\{v\}$
        \Repeat
            \State $u\gets $ highest-numbered vertex in $N$
            \State $N\gets N\cap N_G(u)$\algorithmiccomment{\emph{intersect-gt}, $\theta=|C^*|-|C|$}
            \State $C\gets C\cup \{u\}$
        \Until{$|N|==0$}
        \If{$|C|>|C^*|$} \State $C^*\gets C$
        \EndIf
    \EndParFor
    \EndFunction
  \end{algorithmic}
\end{algorithm}%
Algorithms~\ref{algo:dheur} and~\ref{algo:cheur} describe the
degree-based and coreness-based heurstic search procedures.
The key difference between these algorithms is that degree-based heuristic
search searches for the vertex with highest degree in the subgraph
induced by the candidate set $N$, whereas coreness-based search
simply takes the highest-numbered vertex in $N$ as this will have the
highest coreness due to relabelling.
Both heuristic searches utilize the early-exit set intersection algorithms.

\subsection{Systematic Search}
\begin{algorithm}[t]
  \caption{Systematic search of a neighborhood.}
  \label{algo:sys}
  \begin{algorithmic}[1]
    \small
    \Function{SystematicSearch}{$H$,$o$}
    \ParFor{$k\gets |C^*|$ to $degeneracy+1$}
        \State $v\gets $ lowest-numbered vertex $u$ with $c[v]=k$
        \State NeighborSearch($H$,$v$)
    \EndParFor
    \For{$k\gets degeneracy+1$ downto $1$}
        \If{$k\ge |C^*|$}
            \ParFor{$v\in V: c[v]=k$}
                \If{$c[v]\ge |C^*|$} \State NeighborSearch($H$,$v$)
                \EndIf
            \EndParFor
        \EndIf
    \EndFor
    \EndFunction
  \end{algorithmic}
\end{algorithm}
\begin{algorithm}[t]
  \caption{Searching the right-neighborhood of a vertex.}
  \label{algo:ngh}
  \begin{algorithmic}[1]
    \small
    \Function{NeighborSearch}{$H$,$v$}
        \State $N\gets \{u\in N_G^{+}(v): c[u]\ge |C^*|\}$ \algorithmiccomment{filter 1}
        \IIf{$|N|<|C^*|$} \Return \EndIIf
        \For{$u\in N$} \algorithmiccomment{filter 2}
            \If{not \emph{intersect-size-gt-bool}($N_G(u)$,$N$,$|C^*|-2$)}
                \State remove $u$ from $N$
            \EndIf
        \EndFor
        \IIf{$|N|<|C^*|$} \Return \EndIIf
        \State $\hat{m}\gets 0$ \algorithmiccomment{estimated number of edges}
        \For{$u\in N$} \algorithmiccomment{filter 3}
            \State $d\gets$ \emph{intersect-size-gt-val}($N_G(u)$,$N$,$|C^*|-2$)
            \State remove $u$ from $N$ if $d\le |C^*|-2$
            \State $\hat{m}\gets \hat{m} + d$
        \EndFor
        \IIf{$|N|<|C^*|$} \Return \EndIIf
        \If{$\hat{m}/|V|/(|V|-1) > \phi$} \algorithmiccomment{$0\le\phi\le 1$, tunable}
            \State k-VertexCover($G[N]$,$|N|+1-|C^*|$)
        \Else
            \State MaximumCliqueSearch($G[N]$,$\{v\}$)
        \EndIf
    \EndFunction
  \end{algorithmic}
\end{algorithm}
The key task of systematic search is to establish the maximum clique
by systematically considering every eligible vertex and its
right neighborhood (Algorithm~\ref{algo:sys}).
We first consider one random vertex per degeneracy level, followed
by searching all degeneracy levels from high to low and
processing all vertices with the same degeneracy in parallel.
Imposing an order from high to low degeneracy levels links in with the
distinction between \emph{must} and \emph{may} vertices:
first the \emph{must} vertices are
explored, followed by \emph{may} vertices.
Searching some randomly selected vertices with low degeneracy
at first improves performance
especially for those graphs with a high clique-core gap, as it helps
to establish a good-sized incumbent clique quickly. The performance
overhead is limited due to analyzing few vertices, and
because lower-degeneracy vertices can typically be evaluated in less time.

For each vertex in the graph, algorithm \emph{NeighborSearch} searches its
right-neighborhood (Algorithm~\ref{algo:ngh}).
Most neighborhoods do not contain a maximum clique.
As such, our approach is optimized to prove absence of a maximum clique
by filtering out many candidate vertices from the top-level neighborhood.
Before starting a search, it filters out elements that are certain
to not contribute to a maximum clique by
(i)~selecting only vertices with coreness at least $|C|$;
(ii)~removing any vertex whose degree in the subgraph induced by the
eligible vertices is insufficient to belong to a clique larger than $|C|$;
(iii)~removing again any vertices that fail to meet the same criterion.
Detailed search is initiated only if the remaining neighborhood
is sufficiently large.
The filtering could be repeated until no further vertices are removed.
However, we find that two iterations of degree-based filtering are
sufficient to exclude search for the majority of neighborhoods.

The threshold $|C|-2$ comes about by requiring strictly more
than $|C|-2$ neighbors to vertex $u$, and adding vertices $u$ and $v$.
Note that $v$ is not included in $N_G^{+}(v)$.

\subsection{Algorithmic Choice}
High-density subgraphs are solved using the $k$-Vertex Cover ($k$-VC) problem
instead of searching direct for a maximum clique.
Similar to dOmega~\cite{walteros:20:easy}, we use repeated $k$-Vertex Cover
calls to find the maximum clique using a binary search over a range of
plausible $k$ values. Differently from dOmega, the binary search applies
to a single neighborhood.
Making a precise prediction of what algorithm is most efficient
is challenging as in general the hardness
of an instance is not easily assessed.
A simple threshold-based assessment based on the density of the
subgraph is, however, already efficient.

We use MC and $k$-VC solvers that implement the key established techniques.
As such they are fairly basic algorithms.
The MC solver is derived from the Bron-Kerbosch algorithm
for Maximal Clique Enumeration~\cite{bk:73:mce} and uses
Tomita's pivoting technique~\cite{tomita:06:pivot} to minimize the
branching factor of the tree.
Vertices are sorted by degeneracy order~\cite{eppstein:10:mce},
which minimizes the size of the right-neighborhoods.
Pruning is performed by comparison of the current clique and candidate
set to the incumbent clique size. A pruning rule based on coloring the
graph induced by the candidate set is also used to obtain an upper bound
on any clique contained in it~\cite{babel:90:mc}.

The $k$-VC solver is a branch-and-bound solver that
branches on the highest-degree vertices.
It resorts to a polynomial time algorithm for paths and cycles when
the maximum degree becomes two.
It implements the Buss kernel~\cite{buss:93:nondet}
that addresses high-degree vertices,
and applies rules for kernelisation of vertices
up to degree 2~\cite{abu:17:kernel,fellows:18:kernel} (although only
those cases are implemented where no vertices are merged).
Kernelisation replaces low-degree vertices such that an equivalent problem
of lower dimensionality is obtained.
This implementation is comparable to the one used
in dOmega~\cite{walteros:20:easy}.

\subsection{Vertex Ordering}
Vertex ordering is highly impactful. It determines the composition
of right-neighborhoods, and as such determines which subproblems are solved,
and it determines the order in which the subproblems are solved.
It has been shown that sorting by increasing degeneracy order
is highly effective~\cite{eppstein:10:mce,ostergard:02:cliquer,carraghan:90:mc}.
In particular, the peeling order, which is the order in which the coreness
of a vertex is determined by
Matula and Beck's algorithm~\cite{matula:83:degeneracy},
guarantees that every right-neighborhood is no larger than the coreness
of the vertex. This ensures that subproblems are relatively small.

To the best of our knowledge,
all MC algorithms applying the degeneracy peeling order
are sequential algorithms.
No uniquely defined peeling order is available
when we calculate coreness using a parallel algorithm.
As such, we sort vertices by increasing coreness
and break ties by sorting by increasing degree.
This sort is performed efficiently in parallel in two phases,
using the SAPCo sort algorithm for sorting by degree~\cite{koohi:22:sapco}
followed by a stable counting sort.

%% file: eval.tex
\section{Experimental Evaluation}
\label{sec:eval}
\input{tab_graphs}
We implemented LazyMC in C++23 using gcc version 13.0. Parallelism is
controlled using Parlay\footnote{\url{https://github.com/cmuparlay/parlaylib}}.
Hash sets are implemented with hopscotch hashing~\cite{herlihy:08:hopscotch}
with maximum distance between the location of a vertex and its home index
equal to the cache line size.
The cache line size is 64 bytes and holds 16 4-byte vertex identifiers.
We use bitmasks to identify elements mapped to the
same home index rather than deltas, as this provided better experimental
performance.

Measurements are conducted on a dual socket Intel Xeon Gold 6438Y+
(Sapphire Rapids) with 32 cores per socket and two
hyperthreads per core.
Experiments are performed on a set of 28 large graphs,
which are detailed in Table~\ref{tab:graphs}.

\input{tab_overall}

\subsection{Overall Performance}
We compare the overall runtime of LazyMC to state-of-the-art
algorithms.
PMC~\cite{rossi:14:pmc} is a parallel MC algorithm that uses
coreness-based heuristic search and prunes the search tree using
graph coloring.
dOmega~\cite{walteros:20:easy} 
performs multiple searches over the graph, each aiming
to find a clique with increasingly higher clique-core gap.
Neighborhoods are sorted by coreness and are solved using the
$k$-vertex cover problem where $k$ is determined by the clique-core gap
currently searched for. In cases where the clique-core gap is not zero,
it is useful to distinguish a linear progression (LS)
of clique-core gap (0, 1, 2, $\ldots$), and a binary search (BS).
The higher end of the range for the clique-core gap is given by
the degeneracy order heuristic. dOmega is a sequential algorithm.
MC-BRB~\cite{chang:19:brb} is a sequential branch-and-bound algorithm
that applies extensive filtering at each branching step.
The filtering rules
are an extension of~\cite{fahle:02:mc} and allow to remove or
fold vertices in the presense of high-degree vertices in the subgraph
induced by the candidate set $(G[P])$.
MC-BRB furthermore uses a degree-based heuristic search.
We present results on sequential algorithms primarily because the focus
in the field is on algorithmic innovation and parallel codes are rare.
Note that all algorithms compute the exact maximum clique.

Table~\ref{tab:overall} summarizes the execution time and
standard deviation (as a percentage of execution time) for the five algorithms,
and shows the speedup of LazyMC over each algorithm.
A timeout of 30 minutes was set for all cases, which includes reading the
graph from disk.
Each algorithm performs exceptionally well on some graphs, and
poor on others. This is to be expected
within the realm of NP-hard algorithms, especially because the algorithms
are constructed in strongly different ways.

LazyMC achieves a median speedup of 3.12 over PMC, and a speedup up to 38.8
(sinaweibo). It is slower only on hollywood.
As LazyMC is most similar to PMC, we attribute this to the key
contributions of LazyMC: lazy graph construction, early exit intersections
and the use of vertex cover to solve subproblems.

LazyMC is faster than dOmega, with a median speedup of 6.23 over LS,
and 3.87 over BS. The difference between the LS and BS variants especially
where the clique-core gap is high
has been noted before~\cite{walteros:20:easy}.

LazyMC outperforms also MC-BRB with a median speedup of 2.38. MC-BRB is however faster on a number of graphs. These include small graphs with zero clique-core gap
(dblp, it, hollywood, uk) and graphs with non-zero but relatively small
clique-core gap (HS-CX, mouse). We will discuss later the reasons for this.
On the flickr graph, MC-BRB's filtering rules work exceptionally well.
These rules could be easily added to LazyMC.
Note that PMC and dOmega-LS do not complete flickr in the alloted time.

\subsection{Analysis}
\begin{figure}[!t]
  \includegraphics[width=.99\columnwidth]{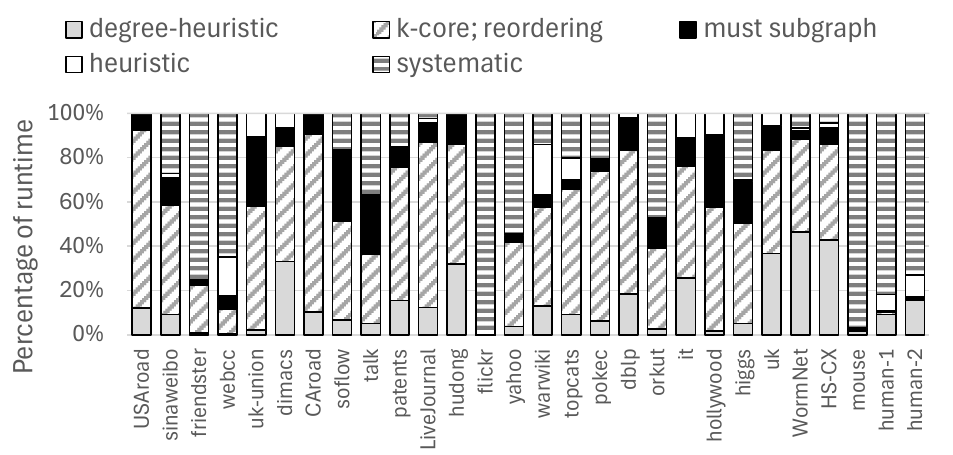}
  \caption{Relative time spent in the key steps of LazyMC.}
  \label{fig:r:steps}
\end{figure}%
We analyze the relative amount of time spent in the key steps of the algorithm
(Figure~\ref{fig:r:steps}).
For many graphs, time is dominated by the k-core computation
and determining the vertex order.
This includes some of the small graphs where LazyMC is outperformed by MC-BRB.
Here, the sorting weighs heavily on execution time without providing substantial
filtering benefits.
In contrast,
MC-BRB obtains the sort order for free by using a sequential k-core algorithm.

In some algorithms substantial time is spent constructing
a reordered and hashed representation of the \emph{must} subgraph.
Note that this subgraph is determined based on knowledge of the incumbent
clique found by degree-based heuristic search. As such, sometimes more time
is spent pre-populating the hashed representation than what is useful.
There is scope to fine-tune the fraction of the graph that should be
non-lazily constructed. We found however that constructing the graph entirely
in a lazy fashion generally produces slower execution times
than those presented.

\begin{figure}[!t]
  \includegraphics[width=.99\columnwidth]{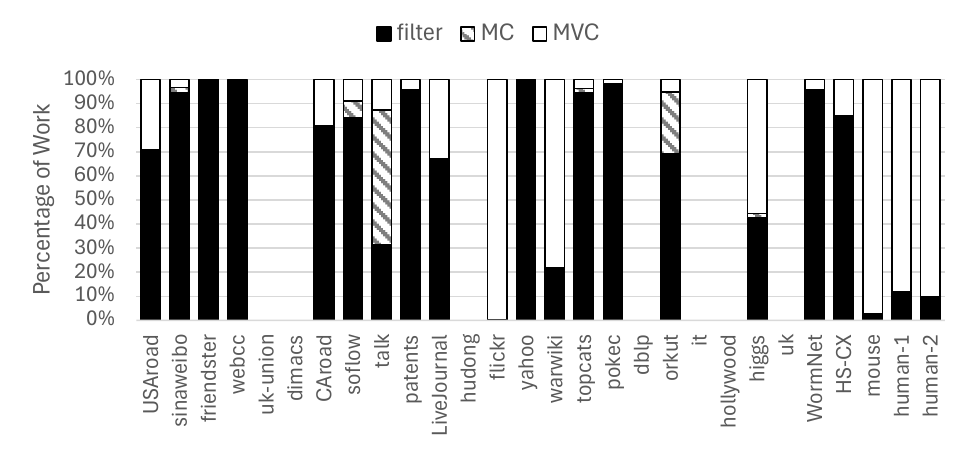}
  \caption{Break-down of time spent in systematic search.
    Where graphs have no data, a maximum clique with clique-core gap zero
    is found during heuristic search.}
  \label{fig:r:algo}
\end{figure}

The time spent during systematic search is depicted in Figure~\ref{fig:r:algo}.
We observe that the vertex cover algorithm is predominantly selected
to search subgraphs. This is due to selecting vertex cover
when the density of the subgraph is 50\% or higher.
The impact of this threshold is analysed in Section~\ref{sec:eval:choice}.

\input{tab_filter}

Filtering prior to searching a neighborhood
takes up the majority of time in many graphs. In some instances,
e.g., webcc and yahoo, it is also
successful to disprove the need to search any neighborhoods at all.
Multiple rounds of filtering are highly important, as removing vertices
can trigger other vertices to be removed too.
Table~\ref{tab:filter} shows the efficacy of the filtering steps.
A first filter considers neighborhoods only when the coreness of the
corresponding vertex is sufficiently large in light of the
incumbent clique size.
Removing the neighbors with insufficient coreness themselves (filter 1)
reduces the neighborhood size sufficiently to discard the neighborhood
in a few cases.
Considering the vertex degree inside the subgraph induced by the neighborhood
(filter 2) is a strong filter and massively reduces the number of instances
that need to be searched. Repeating this filter a second time (filter 2)
reduces the number of neighborhoods to search to a few in a thousand,
except in the dense gene networks. % (mouse, human-1, human-2).

\subsection{Laziness}
\begin{figure}[!t]
  \includegraphics[width=.99\columnwidth]{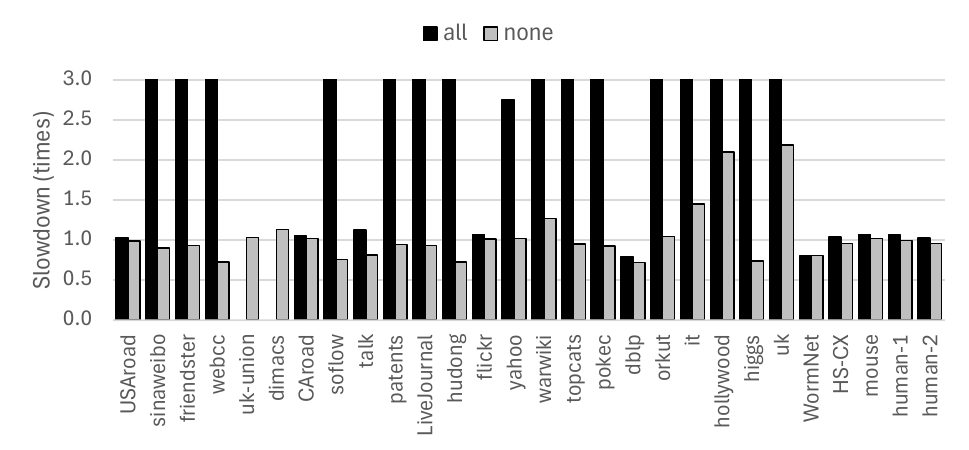}
  \caption{Slowdown incurred when pre-populating all vertices' neighborhoods
    or none. Baseline (1) is pre-populating with \emph{must} subgraph only.
  Graphs uk-union and dimacs incurred out-of-memory errors for case ``all''.}
  \label{fig:r:lazy}
\end{figure}

The lazily constructed hashed graph has a significant performance impact
as it is used extensively during coreness-based heuristic search and
filtering.
In the baseline algorithm we pre-populate all neighborhoods for vertices
in the \emph{must} subgraph, i.e., those with coreness as large as
the incumbent clique size found by degree-based heuristic search.
Figure~\ref{fig:r:lazy} shows the performance impact when pre-populating
all neighborhoods, or pre-populating none. Pre-populating all neighborhoods
is clearly ineffective and major slowdowns are incurred, up to 26$\times$
for the uk graph. Pre-populating no neighborhoods sometimes improves
performance, often in cases where the heuristic search
finds the maximum clique, however, slowdowns are equally likely.
The geometric mean average here is 0.996, compared to 1 for the baseline.

\subsection{Early-Exit Intersections}
\begin{figure}[!t]
  \includegraphics[width=.99\columnwidth]{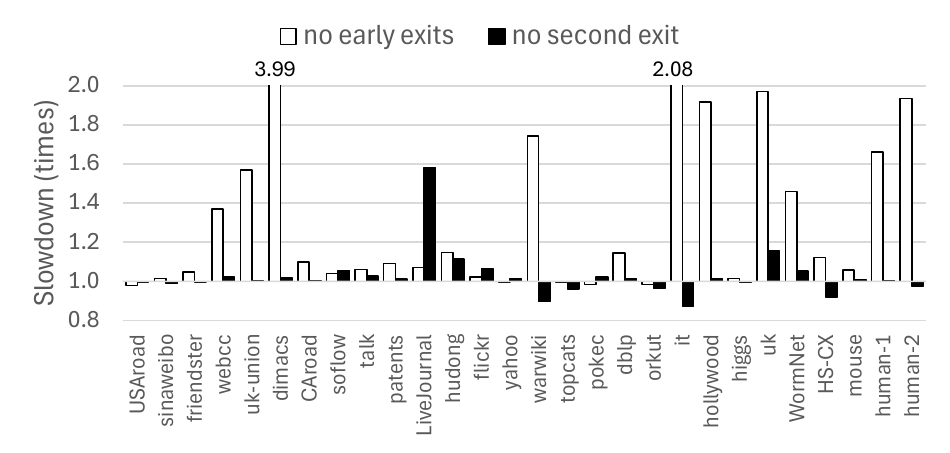}
  \caption{Ablation study on the impact of early-exit intersection
  operations, and the second early exit in \emph{intersect-size-gt-bool}.}
  \label{fig:r:ins}
\end{figure}%
The early exit intersections always improve performance on average over
the execution of the program (Figure~\ref{fig:r:ins}).
The largest performance impact occurs
for dimacs where a 3.99$\times$ slowdown occurs if disabled. The slowdown
is primarily incurred in the degree-based heuristic search,
which uses \emph{intersect-size-gt-val} to determine the highest-degree
vertex in the candidate list. This step is slowed down by about 10$\times$.
Also the heuristic search, which uses \emph{intersect-gt},
is substantially slowed-down by about 4$\times$.

Figure~\ref{fig:r:ins} also shows the impact of the second early exit
in \emph{intersect-size-gt-bool}. This primarily affects filtering.
This second exit is mostly important in LiveJournal and uk.
The second exit does not always improve performance: warwiki and it would
execute about 10\% faster without it.
The negative impact occurs when the operation does not exit early frequently
enough with a \emph{true} outcome as it incurs a small but continuous overhead
that needs to be offset. It should also be observed that this negative impact
is small compared to the variability on the execution time
(Table~\ref{tab:overall}).

\subsection{Algorithmic Choice}
\label{sec:eval:choice}
\begin{figure}[!t]
  \subfloat[Performance]{\label{fig:r:choice:perf}
    \centering
    \includegraphics[width=.40\columnwidth]{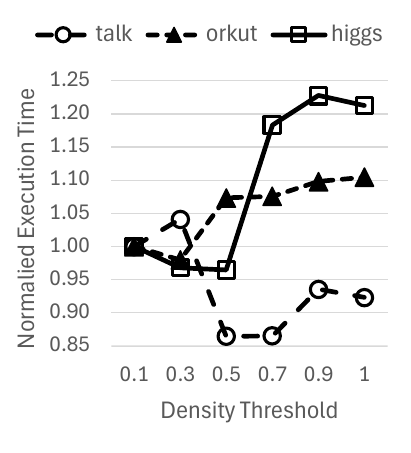}
  }%
  \subfloat[orkut]{\label{fig:r:choice:orkut}
    \centering
    \includegraphics[width=.27\columnwidth]{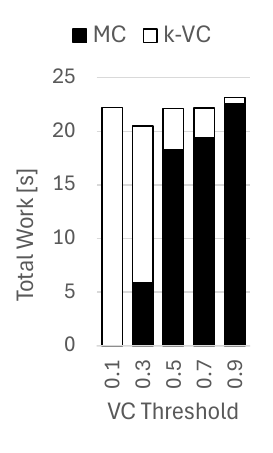}
  }%
  \subfloat[higgs]{\label{fig:r:choice:higgs}
    \centering
    \includegraphics[width=.27\columnwidth]{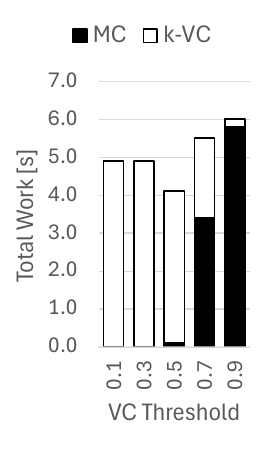}
  }
  \caption{Impact of algorithm choice: Maximum Clique (MC)
    vs.\ $k$-Vertex Cover ($k$-VC).}
  \label{fig:r:choice}
\end{figure}%
\begin{figure*}[!t]
  \centering
  \subfloat[patents]{\label{fig:par:patents}
    \centering
    \includegraphics[width=.45\columnwidth]{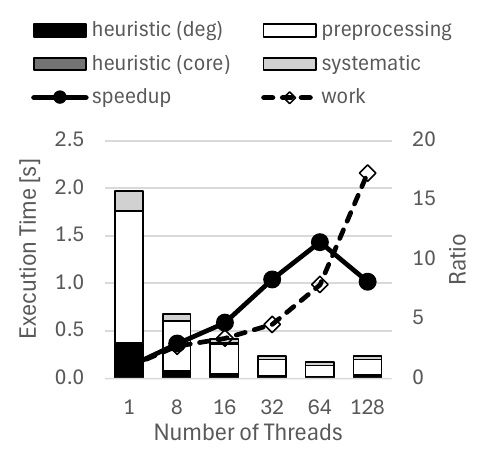}
  }%
  \subfloat[warwiki]{\label{fig:par:warwiki}
    \centering
    \includegraphics[width=.45\columnwidth]{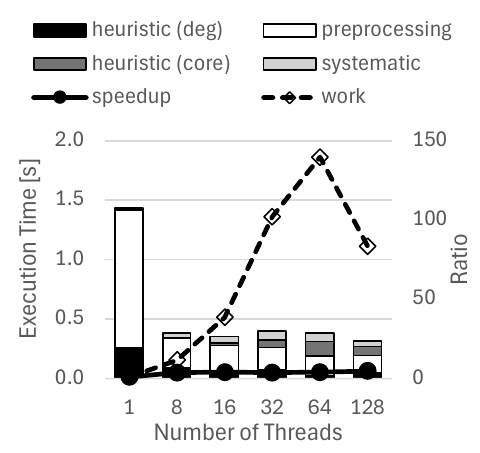}
  }%
  \subfloat[orkut]{\label{fig:par:orkut}
    \centering
    \includegraphics[width=.45\columnwidth]{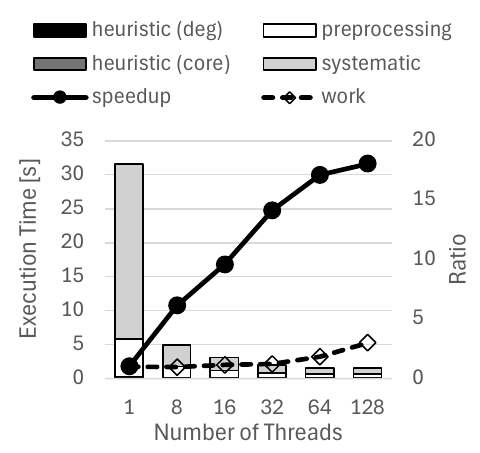}
  }%
  \subfloat[human-1]{\label{fig:par:human1}
    \centering
    \includegraphics[width=.45\columnwidth]{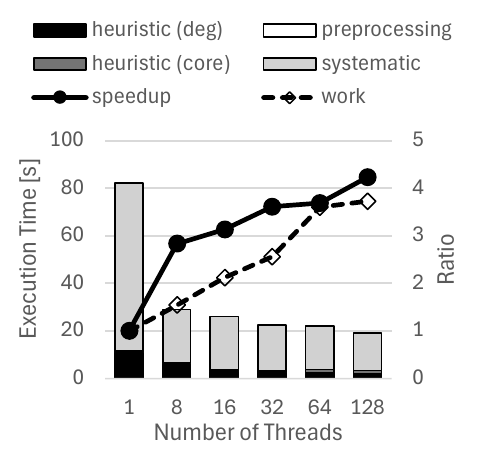}
  }%
  \caption{Efficiency of parallel scaling and adverse impact on total work.}
  \label{fig:par}
\end{figure*}%
The Minimum Vertex Cover often allows to find solutions faster
than searching directly for a Maximum Clique.
For each subgraph is decided independently which algorithm should be used,
based on the density of the subgraph. Densities above a pre-defined
threshold trigger $k$-VC. The main results assume a threhold of 10\%.
Figure~\ref{fig:r:choice:perf} shows that the correct algorithm choice for
each subgraph matters.
E.g., orkut sees a reduction in work by raising the threshold for $k$-VC
from 10\% to 30\%, indicating subgraphs in this range execute on average
faster with MC. Subgraphs with density in the range 30\%--50\% however
are more quickly executed using the $k$-VC.
A similar trend occurs for higgs, but here subgraphs that benefit from
MC sit in the 30\%--50\% range whereas those in the 50\%--70\% are more
efficiently processed with $k$-VC.

\subsection{Parallel Scalability}
Parallel execution is expected to reduce end-to-end execution time,
however, there are several effects that counter this.
Figure~\ref{fig:par} shows the execution time breakdown of the
full algorithm over 4 phases, the speedup, and the amount of work
performed in the systematic search.

Not all phases scale well with an increasing number of threads.
Especially the k-core computation has non-trivial
parallel efficiency~\cite{dhulipala:21:gbbs}.
This is clear in warwiki (Figure~\ref{fig:par:warwiki}) where the
preprocessing stage takes between 0.25 and 0.15 seconds for thread
counts between 8 and 128.

For other steps, parallelism ignores or foregoes improvements found
to the incumbent clique. Where a neighborhood would be searched
more quickly given knowledge of an incumbent clique of a certain size, or may
not be searched at all, parallel execution may initiate the search of
that neighborhood prior to the incumbent clique reaching that size.
As such, more work is performed when executing MC in parallel
(see also~\cite{mccreesh:15:topc} for a discussion of this phenomenon).
This affects systematic search and both heuristic search phases.

In patents (Figure~\ref{fig:par:patents})
and human-1 (Figure~\ref{fig:par:human1}),
increasing thread counts creates speedup.
However, the amount of work executed in the systematic search
grows more or less proportionally.
The amount of work shoot ups in warwiki to 139$\times$ whereas
speedup is 4.7 at most (Figure~\ref{fig:par:warwiki}). Here, both the
number of neighborhoods searched and the time per neighborhood increase.
Orkut is one of the more well-behaved cases:
parallelism increases to 18$\times$ and
the amount of work increases by at most 1.82$\times$
(Figure~\ref{fig:par:orkut}).

The opposite effect, where improving the incumbent clique accelerates
concurrently executing searches, has been observed
in the literature~\cite{mccreesh:13:mc}. One reason we may not be seeing
the effect is that we solve many neighborhoods using $k$-vertex cover,
where the $k$ parameter is determined ahead of the search and is not
updated when the incumbent clique size changes.

Across all graphs, the best parallel speedup is 22.8$\times$ on 128 threads
and the worst increase in work is as seen in warwiki.
Speedup drops below 1 in just one instance: HS-CX on 128 threads.
Here, the single-thread execution time is just 31.8$\:$ms, leaving little
room for speedup.

%% file: tab_graphs.tex
\begin{table}[t]
  \caption{Characterization of graphs used in experiments,
    showing number of vertices ($|V|$) and edges ($|E|$),
    the maximum degree ($\Delta$), degeneracy ($d$),
    maximum clique size ($\omega$), clique-core gap ($g=d+1-\omega$),
    incumbent clique size found by degree-based heuristic search ($\hat{\omega}_d$),
    and incumbent clique size found by coreness-based heuristic search ($\hat{\omega}_h$). Entries in bold indicate clique-core gap zero, and cases where a heuristic search finds the maximum clique.}
  \label{tab:graphs}
  \centering
  \footnotesize
  \setlength\tabcolsep{1pt}
  \begin{tabular}{|l|r|r|r|r|r|r|r|r|r|}
\hline
& $|V|$ & $|E|$ & $\Delta$ & $d$ & $\omega$ & $g$ & $\hat{\omega}_d$ & $\hat{\omega}_h$ \\
\hline
\textbf{USAroad}~\cite{rossie:15:ndr} &	23.9M &	57.7M &	9 &	3 &	\textbf{4} &	0 &	3 &	3 \\
\textbf{sinaweibo}~\cite{rossie:15:ndr} &	58.7M &	523M &	278K &	193 &	44 &	150 &	8 &	15 \\
\textbf{friendster}~\cite{friendster} &	125M &	5.17G &	5,365 &	269 &	12 &	258 &	3 &	3 \\
\textbf{webcc}~\cite{meusel:14:graph} &	89.1M &	3.87G &	3.0M &	10,487 &	2,935 &	7,553 &	75 &	\textbf{2,935} \\
\textbf{uk-union}-06~\cite{boldi:09:uk} &	132M &	9.33G &	6.4M &	3,628 &	\textbf{3,629} &	0 &	29 &	\textbf{3,629} \\
\textbf{dimacs}~\cite{rossie:15:ndr} &	105M &	6.60G &	975K &	5,704 &	\textbf{5,705} &	0 &	82 &	\textbf{5,705} \\
\textbf{CAroad}~\cite{rossie:15:ndr} &	1.97M &	5.53M &	12 &	3 &	\textbf{4} &	0 &	3 &	3 \\
sx-\textbf{s}tack\textbf{o}ver\textbf{flow}~\cite{SNAP} &	6.02M &	56.4M &	44.1K &	198 &	55 &	144 &	10 &	41 \\
wiki-\textbf{talk}~\cite{SNAP} &	2.39M &	9.32M &	100K &	131 &	26 &	106 &	3 &	20 \\
cit-\textbf{patents}~\cite{SNAP} &	3.77M &	33.0M &	793 &	64 &	11 &	54 &	6 &	6 \\
\textbf{LiveJournal}~\cite{SNAP} &	4.85M &	85.7M &	20.0K &	372 &	321 &	52 &	27 &	307 \\
\textbf{hudong}~\cite{rossie:15:ndr} &	1.98M &	28.9M &	61.4K &	266 &	\textbf{267} &	0 &	245 &	\textbf{267} \\
\textbf{flickr}~\cite{SNAP} &	1.72M &	31.1M &	27.2K &	568 &	98 &	471 &	7 &	70 \\
\textbf{yahoo}-member~\cite{yahoo:webscope} &	1.64M &	30.4M &	5,429 &	49 &	2 &	48 &	\textbf{2} &	2 \\
\textbf{warwiki}~\cite{warwiki} &	2.09M &	52.1M &	1.1M &	893 &	873 &	21 &	243 &	871 \\
wiki-\textbf{topcats}~\cite{rossie:15:ndr} &	1.79M &	50.9M &	238K &	99 &	39 &	61 &	7 &	18 \\
\textbf{pokec}~\cite{kunegis:13:konect} &	1.63M &	44.6M &	14.9K &	47 &	29 &	19 &	18 &	18 \\
\textbf{dblp}2012~\cite{SNAP} &	317K &	2.10M &	343 &	113 &	\textbf{114} &	0 &	18 &	\textbf{114} \\
\textbf{orkut}~\cite{mislove-2007-socialnetworks} &	3.1M &	234M &	33.3K &	253 &	51 &	203 &	27 &	27 \\
\textbf{it}-2004~\cite{boldi:04:cmp} &	509k &	14.4M &	469 &	431 &	\textbf{432} &	0 &	93 &	\textbf{432} \\
\textbf{hollywood}2009~\cite{boldi:04:cmp} &	1.1M &	113M &	11.5K &	2,208 &	\textbf{2,209} &	0 &	66 &	\textbf{2,209} \\
\textbf{higgs}-twitter~\cite{rossie:15:ndr} &	457K &	25.0M &	51.4K &	125 &	71 &	55 &	36 &	36 \\
\textbf{uk}-2005~\cite{rossie:15:ndr} &	130K &	23.5M &	850 &	499 &	\textbf{500} &	0 &	294 &	\textbf{500} \\
bio-\textbf{WormNet}-v3~\cite{rossie:15:ndr} &	16.3K &	1.53M &	1,272 &	164 &	121 &	44 &	119 &	119 \\
bio-\textbf{HS-CX}~\cite{rossie:15:ndr} &	4.41K &	218K &	473 &	98 &	86 &	13 &	86 &	\textbf{86} \\
bio-\textbf{mouse}-gene~\cite{rossie:15:ndr} &	45.1K &	28.9M &	8,031 &	1,045 &	561 &	485 &	\textbf{561} &	561 \\
bio-\textbf{human}-gene\textbf{-1}~\cite{rossie:15:ndr} &	22.3K &	24.6M &	7,938 &	2,047 &	1,335 &	713 &	\textbf{1,335} &	1,335 \\
bio-\textbf{human}-gene\textbf{-2}~\cite{rossie:15:ndr} &	14.3K &	18.1M &	7,228 &	1,902 &	1,300 &	603 &	1,299 &	1,299 \\
\hline
  \end{tabular}
  \end{table}

%% file: tab_overall.tex
\begin{table*}[t]
  \caption{Overall result, showing execution time (PMC and LazyMC use 128 threads),
    standard deviation of execution time as a percentage of the same, and speedup of LazyMC.
    ``T.O.'' indicates timeout. ``Error'' relates to an inability to work with
    more than 4 billion edges.}
  \label{tab:overall}
  \centering
  \footnotesize
  \setlength\tabcolsep{3pt}
  \begin{tabular}{|l||r|r|r||r|r|r||r|r|r||r|r|r||r|r|}
\hline
 & \multicolumn{3}{c||}{\textbf{PMC}}
 & \multicolumn{3}{c||}{\textbf{dOmega-LS}}
 & \multicolumn{3}{c||}{\textbf{dOmega-BS}}
 & \multicolumn{3}{c||}{\textbf{MC-BRB}}
 & \multicolumn{2}{c|}{\textbf{LazyMC}} \\
\hline
& Time & Dev(\%) & Speedup
& Time & Dev(\%) & Speedup
& Time & Dev(\%) & Speedup
& Time & Dev(\%) & Speedup
& Time & Dev(\%) \\
\hline
USAroad &	6.657 &	14.68 &	7.84 &	4.511 &	3.06 &	5.31 &	4.575 &	3.18 &	5.39 &	1.051 &	21.80 &	1.24 &	0.849 &	1.58 \\
sinaweibo &	85.878 &	2.08 &	38.85 &	208.704 &	1.08 &	94.41 &	208.948 &	1.48 &	94.52 &	17.876 &	2.94 &	8.09 &	2.211 &	1.58 \\
friendster &	T.O. &	x &	x &	Error &	x &	x &	Error &	x &	x &	T.O. &	x &	x &	49.978 &	1.03 \\
webcc &	T.O. &	x &	x &	T.O. &	x &	x &	T.O. &	x &	x &	T.O. &	x &	x &	51.777 &	4.29 \\
uk-union &	T.O. &	x &	x &	T.O. &	x &	x &	T.O. &	x &	x &	T.O. &	x &	x &	21.343 &	5.23 \\
dimacs &	45.844 &	44.44 &	3.12 &	Error &	x &	x &	Error &	x &	x &	T.O. &	x &	x &	14.699 &	3.03 \\
CAroad &	0.161 &	3.42 &	1.27 &	0.292 &	6.23 &	2.30 &	0.325 &	12.39 &	2.57 &	0.162 &	19.47 &	1.28 &	0.127 &	8.01 \\
soflow &	10.339 &	4.26 &	20.25 &	42.182 &	4.74 &	82.64 &	43.115 &	1.49 &	84.47 &	4.877 &	18.26 &	9.56 &	0.510 &	4.75 \\
talk &	0.976 &	9.80 &	2.43 &	5.274 &	5.49 &	13.11 &	3.541 &	14.22 &	8.80 &	1.144 &	3.09 &	2.84 &	0.402 &	22.04 \\
patents &	1.683 &	8.56 &	6.46 &	2.236 &	6.07 &	8.58 &	2.132 &	9.54 &	8.19 &	1.207 &	25.25 &	4.63 &	0.260 &	19.29 \\
LiveJournal &	0.826 &	15.95 &	2.33 &	2.399 &	17.94 &	6.78 &	1.799 &	4.89 &	5.08 &	1.232 &	28.42 &	3.48 &	0.354 &	6.12 \\
hudong &	0.411 &	11.41 &	2.98 &	0.496 &	0.69 &	3.60 &	0.533 &	13.35 &	3.87 &	0.616 &	31.48 &	4.46 &	0.138 &	16.71 \\
flickr &	T.O. &	x &	x &	T.O. &	x &	x &	1412.050 &	1.73 &	2.97 &	34.225 &	0.99 &	0.07 &	475.045 &	19.60 \\
yahoo &	2.666 &	5.59 &	7.65 &	12.031 &	57.03 &	34.52 &	12.664 &	53.94 &	36.33 &	2.681 &	22.84 &	7.69 &	0.349 &	4.84 \\
warwiki &	1.896 &	14.84 &	5.65 &	0.511 &	1.23 &	1.52 &	0.396 &	1.08 &	1.18 &	0.716 &	26.78 &	2.13 &	0.335 &	14.70 \\
topcats &	3.719 &	8.21 &	11.87 &	10.595 &	7.07 &	33.83 &	10.813 &	9.78 &	34.52 &	2.329 &	16.86 &	7.44 &	0.313 &	9.21 \\
pokec &	1.679 &	9.89 &	7.82 &	10.022 &	3.61 &	46.65 &	10.482 &	4.26 &	48.79 &	1.826 &	24.71 &	8.50 &	0.215 &	7.20 \\
dblp &	0.084 &	13.31 &	1.77 &	0.072 &	20.99 &	1.50 &	0.049 &	25.76 &	1.03 &	0.020 &	15.72 &	0.43 &	0.048 &	24.02 \\
orkut &	13.021 &	4.91 &	7.34 &	189.173 &	0.52 &	106.64 &	185.938 &	0.49 &	104.82 &	19.660 &	4.16 &	11.08 &	1.774 &	8.81 \\
it &	0.077 &	10.07 &	1.46 &	0.063 &	0.77 &	1.18 &	0.063 &	0.60 &	1.19 &	0.041 &	16.48 &	0.78 &	0.053 &	33.00 \\
hollywood &	1.056 &	57.08 &	0.84 &	0.837 &	2.04 &	0.66 &	0.834 &	1.88 &	0.66 &	0.634 &	27.31 &	0.50 &	1.259 &	18.58 \\
higgs &	1.244 &	9.60 &	2.55 &	11.009 &	8.97 &	22.55 &	13.549 &	5.97 &	27.75 &	2.399 &	17.75 &	4.91 &	0.488 &	3.77 \\
uk &	0.056 &	11.96 &	1.36 &	0.056 &	1.14 &	1.35 &	0.057 &	3.80 &	1.36 &	0.039 &	23.85 &	0.95 &	0.041 &	21.86 \\
WormNet &	0.357 &	19.57 &	6.51 &	1.840 &	27.26 &	33.56 &	1.056 &	20.80 &	19.27 &	0.064 &	2.38 &	1.17 &	0.055 &	1.69 \\
HS-CX &	0.051 &	19.10 &	1.45 &	0.254 &	70.70 &	7.25 &	0.088 &	59.09 &	2.53 &	0.016 &	6.10 &	0.47 &	0.035 &	30.01 \\
mouse &	0.027 &	21.97 &	0.00 &	T.O. &	x &	x &	T.O. &	x &	x &	17.460 &	5.89 &	0.72 &	24.361 &	2.55 \\
human-1 &	T.O. &	x &	x &	146.883 &	3.06 &	7.55 &	16.888 &	1.91 &	0.87 &	45.521 &	2.45 &	2.34 &	19.462 &	1.40 \\
human-2 &	86.392 &	3.42 &	7.47 &	65.854 &	6.45 &	5.69 &	8.932 &	5.18 &	0.77 &	27.328 &	4.23 &	2.36 &	11.571 &	5.86 \\
\hline
median & & & 3.12 & & & 7.40 & & & 5.08 & & & 2.35 & & \\
\hline
  \end{tabular}
\end{table*}

%% file: tab_filter.tex
\begin{table}[t]
  \caption{Fraction of right-neighborhoods retained after each filtering step.
    Normalized per thousand vertices. Graphs where the heuristic
  search finds a zero-gap maximum clique evaluate no right-neighborhoods.}
  \label{tab:filter}
  \centering
  \begin{tabular}{|l||r|r|r|r|}
\hline
 &	coreness &	filter 1 &	filter 2 &	filter 3 \\
\hline
USAroad &	0.007 &	0.001 &	9.6E-05 &	9.6E-05 \\
sinaweibo &	12.171 &	12.095 &	0.074 &	0.041 \\
friendster &	342.610 &	341.772 &	0.035 &	0.012 \\
webcc &	0.770 &	0.559 &	0 &	0 \\
uk-union &	0 &	0 &	0 &	0 \\
dimacs &	0 &	0 &	0 &	0 \\
CAroad &	0.240 &	0.043 &	0.002 &	0.002 \\
soflow &	18.449 &	18.293 &	0.416 &	0.288 \\
talk &	6.220 &	6.135 &	0.769 &	0.656 \\
patents &	38.132 &	26.552 &	0.376 &	0.272 \\
LiveJournal &	0.099 &	0.036 &	0.028 &	0.027 \\
hudong &	0 &	0 &	0 &	0 \\
flickr &	21.978 &	21.819 &	4.905 &	4.478 \\
yahoo &	1000.000 &	982.236 &	0 &	0 \\
warwiki &	0.439 &	0.022 &	0.021 &	0.021 \\
topcats &	56.303 &	53.820 &	0.447 &	0.266 \\
pokec &	160.211 &	136.838 &	0.415 &	0.223 \\
dblp &	0 &	0 &	0 &	0 \\
orkut &	241.365 &	219.664 &	10.940 &	7.682 \\
it &	0 &	0 &	0 &	0 \\
hollywood &	0 &	0 &	0 &	0 \\
higgs &	102.749 &	99.568 &	10.157 &	6.290 \\
uk &	0 &	0 &	0 &	0 \\
WormNet &	130.972 &	116.211 &	5.114 &	2.166 \\
HS-CX &	31.271 &	10.673 &	3.626 &	2.719 \\
mouse &	319.439 &	273.799 &	89.082 &	79.750 \\
human-1 &	203.473 &	142.826 &	115.514 &	113.082 \\
human-2 &	265.481 &	174.038 &	140.990 &	137.789 \\
\hline
  \end{tabular}
\end{table}

%% file: rela.tex
\section{Further Related Work}
\label{sec:rela}
Various works aim to optimize the performance of MC and clique problems
in general.
Hardware bit-level parallelism has been used to accelerate
set intersections~\cite{sansegundo:11:mcbit,dasari:14:pbit}.
Blanusa~\emph{et al.}~\cite{blanusa:20:mce} parallelize
the Bron-Kerbosch~\cite{bk:73:mce} using Intel Threading Building Blocks
and accelerate set intersections using hash intersection.
Besta~\emph{et al.} explore various hash set algorithsm
and compressed bitmaps~\cite{besta:21:gms}.
Vertically vectorized set intersections
and early exit intersections were introduced in~\cite{vandierendonck:24:mce}.
McCreesh and Prosser~\cite{mccreesh:13:mc} observed superlinear speedups
in a multi-threaded MC solver, which they attribute to using parallelism
to avoid commitment to an early heuristic.

Recently attention has been given to optimizing clique problems on GPU.
An early work used the structure of the
Bron-Kerbosch algorithm and used hardware bit parallelism
in the GPU~\cite{vancompernolle:16:bbmcg}.
The depth-first search program structure
is not a good fit to GPU architecture due to the differences in path
length and workload on each of the paths in the search tree.
Geil~\emph{et al.} counter this by using a
breadth-first search~\cite{geil:23:mcegpu}.
Cardone~\emph{et al.} compare between block and warp-level parallelism
and adapt accordingly~\cite{cardone:24:mcgpu}.
Wei~\emph{et al.} apply warp reductions and optimize use
of coalesced memory, in this case for
Maximal Clique Enumeration (MCE)~\cite{wei:21:mcegpu}.
Load imbalance has been addressed with fine-grained queueing
for MCE~\cite{almasri:23:mcegpu} and k-clique counting~\cite{almasri:22:kc-gpu}.

Others consider different, approximate algorithms for the MC problem,
such as neural networks~\cite{cruz:13:mcgpu},
or the approximate Motzkin-Strauss algoritm~\cite{daniluk:19:mcgpu}.

%% file: concl.tex
\section{Conclusion}
\label{sec:concl}
Key to optimization problems like the MC problem is that a significant
part of the graph can be ruled from inspection as soon as some progress
is made towards finding the optimum solution. We leverage this observation
to propose several data structures and algorithms that underpin MC algorithms.
We have implemented these techniques
in an MC algorithm that otherwise uses established search heuristics.
Experimental evaluation demonstrates
performance gains across 28 large graphs with median speedups
over state of the art MC algorithms
of 2.35 (PMC), 5.08 (dOmega) and 2.35 (MC-BRB).
We anticipate that these speedups will be larger when incorporating 
similarly advanced algorithmic ideas as the baselines.

%% file: arxiv.bbl
\begin{thebibliography}{10}
\providecommand{\url}[1]{#1}
\csname url@samestyle\endcsname
\providecommand{\newblock}{\relax}
\providecommand{\bibinfo}[2]{#2}
\providecommand{\BIBentrySTDinterwordspacing}{\spaceskip=0pt\relax}
\providecommand{\BIBentryALTinterwordstretchfactor}{4}
\providecommand{\BIBentryALTinterwordspacing}{\spaceskip=\fontdimen2\font plus
\BIBentryALTinterwordstretchfactor\fontdimen3\font minus
  \fontdimen4\font\relax}
\providecommand{\BIBforeignlanguage}[2]{{%
\expandafter\ifx\csname l@#1\endcsname\relax
\typeout{** WARNING: IEEEtran.bst: No hyphenation pattern has been}%
\typeout{** loaded for the language `#1'. Using the pattern for}%
\typeout{** the default language instead.}%
\else
\language=\csname l@#1\endcsname
\fi
#2}}
\providecommand{\BIBdecl}{\relax}
\BIBdecl

\bibitem{matula:83:degeneracy}
\BIBentryALTinterwordspacing
D.~W. Matula and L.~L. Beck, ``Smallest-last ordering and clustering and graph
  coloring algorithms,'' \emph{J. ACM}, vol.~30, no.~3, p. 417–427, Jul.
  1983. [Online]. Available: \url{https://doi.org/10.1145/2402.322385}
\BIBentrySTDinterwordspacing

\bibitem{blanusa:20:mce}
\BIBentryALTinterwordspacing
J.~Blanu\v{s}a, R.~Stoica, P.~Ienne, and K.~Atasu, ``Manycore clique
  enumeration with fast set intersections,'' \emph{Proc. VLDB Endow.}, vol.~13,
  no.~12, p. 2676–2690, jul 2020. [Online]. Available:
  \url{https://doi.org/10.14778/3407790.3407853}
\BIBentrySTDinterwordspacing

\bibitem{besta:21:gms}
\BIBentryALTinterwordspacing
M.~Besta, Z.~Vonarburg-Shmaria, Y.~Schaffner, L.~Schwarz, G.~Kwasniewski,
  L.~Gianinazzi, J.~Beranek, K.~Janda, T.~Holenstein, S.~Leisinger,
  P.~Tatkowski, E.~Ozdemir, A.~Balla, M.~Copik, P.~Lindenberger, M.~Konieczny,
  O.~Mutlu, and T.~Hoefler, ``Graphminesuite: Enabling high-performance and
  programmable graph mining algorithms with set algebra,'' \emph{Proc. VLDB
  Endow.}, vol.~14, no.~11, p. 1922–1935, jul 2021. [Online]. Available:
  \url{https://doi.org/10.14778/3476249.3476252}
\BIBentrySTDinterwordspacing

\bibitem{vandierendonck:24:mce}
\BIBentryALTinterwordspacing
H.~Vandierendonck, ``Differentiating set intersections in maximal clique
  enumeration by function and subproblem size,'' in \emph{Proceedings of the
  38th ACM International Conference on Supercomputing}, ser. ICS '24.\hskip 1em
  plus 0.5em minus 0.4em\relax New York, NY, USA: Association for Computing
  Machinery, 2024, p. 150–163. [Online]. Available:
  \url{https://doi.org/10.1145/3650200.3656607}
\BIBentrySTDinterwordspacing

\bibitem{besta:21:sisa}
\BIBentryALTinterwordspacing
M.~Besta, R.~Kanakagiri, G.~Kwasniewski, R.~Ausavarungnirun, J.~Ber\'{a}nek,
  K.~Kanellopoulos, K.~Janda, Z.~Vonarburg-Shmaria, L.~Gianinazzi, I.~Stefan,
  J.~G. Luna, J.~Golinowski, M.~Copik, L.~Kapp-Schwoerer, S.~Di~Girolamo,
  N.~Blach, M.~Konieczny, O.~Mutlu, and T.~Hoefler, ``Sisa: Set-centric
  instruction set architecture for graph mining on processing-in-memory
  systems,'' in \emph{MICRO-54: 54th Annual IEEE/ACM International Symposium on
  Microarchitecture}, ser. MICRO '21.\hskip 1em plus 0.5em minus 0.4em\relax
  New York, NY, USA: Association for Computing Machinery, 2021, p. 282–297.
  [Online]. Available: \url{https://doi.org/10.1145/3466752.3480133}
\BIBentrySTDinterwordspacing

\bibitem{rossi:14:pmc}
\BIBentryALTinterwordspacing
R.~A. Rossi, D.~F. Gleich, A.~H. Gebremedhin, and M.~M.~A. Patwary, ``Fast
  maximum clique algorithms for large graphs,'' in \emph{Proceedings of the
  23rd International Conference on World Wide Web}, ser. WWW '14
  Companion.\hskip 1em plus 0.5em minus 0.4em\relax New York, NY, USA:
  Association for Computing Machinery, 2014, p. 365–366. [Online]. Available:
  \url{https://doi.org/10.1145/2567948.2577283}
\BIBentrySTDinterwordspacing

\bibitem{walteros:20:easy}
\BIBentryALTinterwordspacing
J.~L. Walteros and A.~Buchanan, ``Why is maximum clique often easy in
  practice?'' \emph{Operations Research}, vol.~68, no.~6, pp. 1866--1895, 2020.
  [Online]. Available: \url{https://doi.org/10.1287/opre.2019.1970}
\BIBentrySTDinterwordspacing

\bibitem{chang:19:brb}
\BIBentryALTinterwordspacing
L.~Chang, ``Efficient maximum clique computation over large sparse graphs,'' in
  \emph{Proceedings of the 25th ACM SIGKDD International Conference on
  Knowledge Discovery \& Data Mining}, ser. KDD '19.\hskip 1em plus 0.5em minus
  0.4em\relax New York, NY, USA: Association for Computing Machinery, 2019, p.
  529–538. [Online]. Available: \url{https://doi.org/10.1145/3292500.3330986}
\BIBentrySTDinterwordspacing

\bibitem{carraghan:90:mc}
\BIBentryALTinterwordspacing
R.~Carraghan and P.~M. Pardalos, ``An exact algorithm for the maximum clique
  problem,'' \emph{Operations Research Letters}, vol.~9, no.~6, pp. 375--382,
  1990. [Online]. Available:
  \url{https://www.sciencedirect.com/science/article/pii/016763779090057C}
\BIBentrySTDinterwordspacing

\bibitem{tomita:03:mcq}
E.~Tomita and T.~Seki, ``An efficient branch-and-bound algorithm for finding a
  maximum clique,'' in \emph{Discrete Mathematics and Theoretical Computer
  Science}, C.~S. Calude, M.~J. Dinneen, and V.~Vajnovszki, Eds.\hskip 1em plus
  0.5em minus 0.4em\relax Berlin, Heidelberg: Springer Berlin Heidelberg, 2003,
  pp. 278--289.

\bibitem{ostergard:02:cliquer}
\BIBentryALTinterwordspacing
P.~R. Östergård, ``A fast algorithm for the maximum clique problem,''
  \emph{Discrete Applied Mathematics}, vol. 120, no.~1, pp. 197--207, 2002,
  special Issue devoted to the 6th Twente Workshop on Graphs and Combinatorial
  Optimization. [Online]. Available:
  \url{https://www.sciencedirect.com/science/article/pii/S0166218X01002906}
\BIBentrySTDinterwordspacing

\bibitem{lu:17:rmc}
\BIBentryALTinterwordspacing
C.~Lu, J.~X. Yu, H.~Wei, and Y.~Zhang, ``Finding the maximum clique in massive
  graphs,'' \emph{Proc. VLDB Endow.}, vol.~10, no.~11, p. 1538–1549, aug
  2017. [Online]. Available: \url{https://doi.org/10.14778/3137628.3137660}
\BIBentrySTDinterwordspacing

\bibitem{eppstein:10:mce}
\BIBentryALTinterwordspacing
D.~Eppstein, M.~L\"{o}ffler, and D.~Strash, ``{Listing all maximal cliques in
  sparse graphs in near-optimal time},'' in \emph{Exact Complexity of NP-hard
  Problems}, ser. Dagstuhl Seminar Proceedings (DagSemProc), T.~Husfeldt,
  D.~Kratsch, R.~Paturi, and G.~B. Sorkin, Eds., vol. 10441.\hskip 1em plus
  0.5em minus 0.4em\relax Dagstuhl, Germany: Schloss Dagstuhl --
  Leibniz-Zentrum f{\"u}r Informatik, 2011, pp. 1--14. [Online]. Available:
  \url{https://drops.dagstuhl.de/entities/document/10.4230/DagSemProc.10441.2}
\BIBentrySTDinterwordspacing

\bibitem{wu:14:review}
\BIBentryALTinterwordspacing
Q.~Wu and J.-K. Hao, ``A review on algorithms for maximum clique problems,''
  \emph{European Journal of Operational Research}, vol. 242, no.~3, pp.
  693--709, 2015. [Online]. Available:
  \url{https://www.sciencedirect.com/science/article/pii/S0377221714008030}
\BIBentrySTDinterwordspacing

\bibitem{babel:90:mc}
L.~Babel and G.~Tinhofer, ``A branch and bound algorithm for the maximum clique
  problem,'' \emph{Zeitschrift f{\"u}r Operations Research}, vol.~34, pp.
  207--217, 1990.

\bibitem{li:17:momc}
\BIBentryALTinterwordspacing
C.-M. Li, H.~Jiang, and F.~Manyà, ``On minimization of the number of branches
  in branch-and-bound algorithms for the maximum clique problem,''
  \emph{Computers \& Operations Research}, vol.~84, pp. 1--15, 2017. [Online].
  Available:
  \url{https://www.sciencedirect.com/science/article/pii/S0305054817300576}
\BIBentrySTDinterwordspacing

\bibitem{sansegundo:23:clisat}
\BIBentryALTinterwordspacing
P.~{San Segundo}, F.~Furini, D.~Álvarez, and P.~M. Pardalos, ``Clisat: A new
  exact algorithm for hard maximum clique problems,'' \emph{European Journal of
  Operational Research}, vol. 307, no.~3, pp. 1008--1025, 2023. [Online].
  Available:
  \url{https://www.sciencedirect.com/science/article/pii/S0377221722008165}
\BIBentrySTDinterwordspacing

\bibitem{li:10:maxsat}
\BIBentryALTinterwordspacing
C.-M. Li and Z.~Quan, ``An efficient branch-and-bound algorithm based on maxsat
  for the maximum clique problem,'' \emph{Proceedings of the AAAI Conference on
  Artificial Intelligence}, vol.~24, no.~1, pp. 128--133, Jul. 2010. [Online].
  Available: \url{https://ojs.aaai.org/index.php/AAAI/article/view/7536}
\BIBentrySTDinterwordspacing

\bibitem{tomita:10:mcs}
\BIBentryALTinterwordspacing
E.~Tomita, Y.~Sutani, T.~Higashi, S.~Takahashi, and M.~Wakatsuki, ``A simple
  and faster branch-and-bound algorithm for finding a maximum clique,'' in
  \emph{Proceedings of the 4th International Conference on Algorithms and
  Computation}, ser. WALCOM'10.\hskip 1em plus 0.5em minus 0.4em\relax Berlin,
  Heidelberg: Springer-Verlag, 2010, p. 191–203. [Online]. Available:
  \url{https://doi.org/10.1007/978-3-642-11440-3_18}
\BIBentrySTDinterwordspacing

\bibitem{bk:73:mce}
\BIBentryALTinterwordspacing
C.~Bron and J.~Kerbosch, ``Algorithm 457: Finding all cliques of an undirected
  graph,'' \emph{Commun. ACM}, vol.~16, no.~9, p. 575–577, sep 1973.
  [Online]. Available: \url{https://doi.org/10.1145/362342.362367}
\BIBentrySTDinterwordspacing

\bibitem{tomita:06:pivot}
\BIBentryALTinterwordspacing
E.~Tomita, A.~Tanaka, and H.~Takahashi, ``The worst-case time complexity for
  generating all maximal cliques and computational experiments,''
  \emph{Theoretical Computer Science}, vol. 363, no.~1, pp. 28--42, 2006,
  computing and Combinatorics. [Online]. Available:
  \url{https://www.sciencedirect.com/science/article/pii/S0304397506003586}
\BIBentrySTDinterwordspacing

\bibitem{buss:93:nondet}
\BIBentryALTinterwordspacing
J.~F. Buss and J.~Goldsmith, ``Nondeterminism within $p^*$,'' \emph{SIAM
  Journal on Computing}, vol.~22, no.~3, pp. 560--572, 1993. [Online].
  Available: \url{https://doi.org/10.1137/0222038}
\BIBentrySTDinterwordspacing

\bibitem{abu:17:kernel}
F.~N. Abu-Khzam, R.~L. Collins, M.~R. Fellows, M.~A. Langston, W.~H. Suters,
  and C.~T. Symons, ``Kernelization algorithms for the vertex cover problem,''
  2017.

\bibitem{fellows:18:kernel}
\BIBentryALTinterwordspacing
M.~R. Fellows, L.~Jaffke, A.~I. Kir{\'a}ly, F.~A. Rosamond, and M.~Weller,
  \emph{What Is Known About Vertex Cover Kernelization?}\hskip 1em plus 0.5em
  minus 0.4em\relax Cham: Springer International Publishing, 2018, pp.
  330--356. [Online]. Available:
  \url{https://doi.org/10.1007/978-3-319-98355-4\_19}
\BIBentrySTDinterwordspacing

\bibitem{koohi:22:sapco}
M.~Koohi~Esfahani, P.~Kilpatrick, and H.~Vandierendonck, ``{SAPCo} sort:
  Optimizing degree-ordering for power-law graphs,'' in \emph{2022 IEEE
  International Symposium on Performance Analysis of Systems and Software
  (ISPASS)}, 2022, pp. 138--140.

\bibitem{rossie:15:ndr}
\BIBentryALTinterwordspacing
R.~Rossi and N.~Ahmed, ``The network data repository with interactive graph
  analytics and visualization,'' in \emph{Proceedings of the {AAAI} Conference
  on Artificial Intelligence}, vol.~29, no.~1, Mar. 2015. [Online]. Available:
  \url{https://ojs.aaai.org/index.php/AAAI/article/view/9277}
\BIBentrySTDinterwordspacing

\bibitem{friendster}
``Friendster data set,''
  \url{https://archive.org/download/friendster-dataset-201107}, Jul. 2011.

\bibitem{meusel:14:graph}
\BIBentryALTinterwordspacing
R.~Meusel, S.~Vigna, O.~Lehmberg, and C.~Bizer, ``Graph structure in the web
  --- revisited: A trick of the heavy tail,'' in \emph{Proceedings of the 23rd
  International Conference on World Wide Web}, ser. WWW '14 Companion.\hskip
  1em plus 0.5em minus 0.4em\relax New York, NY, USA: Association for Computing
  Machinery, 2014, p. 427–432. [Online]. Available:
  \url{https://doi.org/10.1145/2567948.2576928}
\BIBentrySTDinterwordspacing

\bibitem{boldi:09:uk}
\BIBentryALTinterwordspacing
P.~Boldi, M.~Santini, and S.~Vigna, ``A large time-aware web graph,''
  \emph{SIGIR Forum}, vol.~42, no.~2, p. 33–38, nov 2008. [Online].
  Available: \url{https://doi.org/10.1145/1480506.1480511}
\BIBentrySTDinterwordspacing

\bibitem{SNAP}
\BIBentryALTinterwordspacing
Stanford. (2009) Stanford large network dataset collection. [Online].
  Available: \url{https://snap.stanford.edu/data/}
\BIBentrySTDinterwordspacing

\bibitem{yahoo:webscope}
``Yahoo webscope datasets,'' \url{https://webscope.sandbox.yahoo.com}, last
  retrieved November 2023.

\bibitem{warwiki}
``Wikimedia downloads,'' \url{https://dumps.wikimedia.org}, Jul. 2011.

\bibitem{kunegis:13:konect}
\BIBentryALTinterwordspacing
J.~Kunegis, ``Konect: The koblenz network collection,'' in \emph{Proceedings of
  the 22nd International Conference on World Wide Web}, ser. WWW ’13
  Companion.\hskip 1em plus 0.5em minus 0.4em\relax New York, NY, USA:
  Association for Computing Machinery, 2013, p. 1343–1350. [Online].
  Available: \url{https://doi.org/10.1145/2487788.2488173}
\BIBentrySTDinterwordspacing

\bibitem{mislove-2007-socialnetworks}
\BIBentryALTinterwordspacing
A.~Mislove, M.~Marcon, K.~P. Gummadi, P.~Druschel, and B.~Bhattacharjee,
  ``Measurement and analysis of online social networks,'' in \emph{Proceedings
  of the 7th ACM SIGCOMM Conference on Internet Measurement}, ser. IMC
  '07.\hskip 1em plus 0.5em minus 0.4em\relax New York, NY, USA: Association
  for Computing Machinery, 2007, p. 29–42. [Online]. Available:
  \url{https://doi.org/10.1145/1298306.1298311}
\BIBentrySTDinterwordspacing

\bibitem{boldi:04:cmp}
\BIBentryALTinterwordspacing
P.~Boldi and S.~Vigna, ``The webgraph framework i: Compression techniques,'' in
  \emph{Proceedings of the 13th International Conference on World Wide Web},
  ser. WWW ’04.\hskip 1em plus 0.5em minus 0.4em\relax New York, NY, USA:
  Association for Computing Machinery, 2004, p. 595–602. [Online]. Available:
  \url{https://doi.org/10.1145/988672.988752}
\BIBentrySTDinterwordspacing

\bibitem{herlihy:08:hopscotch}
\BIBentryALTinterwordspacing
M.~Herlihy, N.~Shavit, and M.~Tzafrir, ``Hopscotch hashing,'' in
  \emph{Proceedings of the 22nd International Symposium on Distributed
  Computing}, ser. DISC '08.\hskip 1em plus 0.5em minus 0.4em\relax Berlin,
  Heidelberg: Springer-Verlag, 2008, p. 350–364. [Online]. Available:
  \url{https://doi.org/10.1007/978-3-540-87779-0_24}
\BIBentrySTDinterwordspacing

\bibitem{fahle:02:mc}
T.~Fahle, ``Simple and fast: Improving a branch-and-bound algorithm for maximum
  clique,'' in \emph{Proceedings of the 10th Annual European Symposium on
  Algorithms}, ser. ESA '02.\hskip 1em plus 0.5em minus 0.4em\relax Berlin,
  Heidelberg: Springer-Verlag, 2002, p. 485–498.

\bibitem{dhulipala:21:gbbs}
\BIBentryALTinterwordspacing
L.~Dhulipala, G.~E. Blelloch, and J.~Shun, ``Theoretically efficient parallel
  graph algorithms can be fast and scalable,'' \emph{ACM Trans. Parallel
  Comput.}, vol.~8, no.~1, apr 2021. [Online]. Available:
  \url{https://doi.org/10.1145/3434393}
\BIBentrySTDinterwordspacing

\bibitem{mccreesh:15:topc}
\BIBentryALTinterwordspacing
C.~McCreesh and P.~Prosser, ``The shape of the search tree for the maximum
  clique problem and the implications for parallel branch and bound,''
  \emph{ACM Trans. Parallel Comput.}, vol.~2, no.~1, Apr. 2015. [Online].
  Available: \url{https://doi.org/10.1145/2742359}
\BIBentrySTDinterwordspacing

\bibitem{mccreesh:13:mc}
\BIBentryALTinterwordspacing
------, ``Multi-threading a state-of-the-art maximum clique algorithm,''
  \emph{Algorithms}, vol.~6, pp. 618--635, 2013. [Online]. Available:
  \url{https://api.semanticscholar.org/CorpusID:33465883}
\BIBentrySTDinterwordspacing

\bibitem{sansegundo:11:mcbit}
P.~San~Segundo, D.~Rodr{\'\i}guez-Losada, and A.~Jim{\'e}nez, ``An exact
  bit-parallel algorithm for the maximum clique problem,'' \emph{Computers \&
  Operations Research}, vol.~38, no.~2, pp. 571--581, 2011.

\bibitem{dasari:14:pbit}
N.~S. Dasari, R.~Desh, and Z.~M, ``pbitmce: A bit-based approach for maximal
  clique enumeration on multicore processors,'' in \emph{2014 20th IEEE
  International Conference on Parallel and Distributed Systems (ICPADS)}, 2014,
  pp. 478--485.

\bibitem{vancompernolle:16:bbmcg}
M.~VanCompernolle, L.~Barford, and F.~Harris, ``Maximum clique solver using
  bitsets on gpus,'' in \emph{Information Technology: New Generations: 13th
  International Conference on Information Technology}.\hskip 1em plus 0.5em
  minus 0.4em\relax Springer, 2016, pp. 327--337.

\bibitem{geil:23:mcegpu}
A.~Geil, S.~D. Porumbescu, and J.~D. Owens, ``Maximum clique enumeration on the
  gpu,'' in \emph{2023 IEEE International Parallel and Distributed Processing
  Symposium Workshops (IPDPSW)}, 2023, pp. 234--244.

\bibitem{cardone:24:mcgpu}
L.~Cardone, S.~D. Martino, and S.~Quer, ``Efficiently computing maximum clique
  of sparse graphs with many-core graphical processing units,'' in
  \emph{Proceedings of the 19th International Conference on Software
  Technologies - Volume 1: ICSOFT}, INSTICC.\hskip 1em plus 0.5em minus
  0.4em\relax SciTePress, 2024, pp. 539--546.

\bibitem{wei:21:mcegpu}
Y.-W. Wei, W.-M. Chen, and H.-H. Tsai, ``Accelerating the bron-kerbosch
  algorithm for maximal clique enumeration using gpus,'' \emph{IEEE
  Transactions on Parallel and Distributed Systems}, vol.~32, no.~9, pp.
  2352--2366, 2021.

\bibitem{almasri:23:mcegpu}
M.~Almasri, Y.-H. Chang, I.~E. Hajj, R.~Nagi, J.~Xiong, and W.-m. Hwu,
  ``Parallelizing maximal clique enumeration on gpus,'' in \emph{2023 32nd
  International Conference on Parallel Architectures and Compilation Techniques
  (PACT)}, 2023, pp. 162--175.

\bibitem{almasri:22:kc-gpu}
\BIBentryALTinterwordspacing
M.~Almasri, I.~E. Hajj, R.~Nagi, J.~Xiong, and W.-m. Hwu, ``Parallel k-clique
  counting on gpus,'' in \emph{Proceedings of the 36th ACM International
  Conference on Supercomputing}, ser. ICS '22.\hskip 1em plus 0.5em minus
  0.4em\relax New York, NY, USA: Association for Computing Machinery, 2022.
  [Online]. Available: \url{https://doi.org/10.1145/3524059.3532382}
\BIBentrySTDinterwordspacing

\bibitem{cruz:13:mcgpu}
R.~Cruz, N.~López, and C.~Trefftz, ``Parallelizing a heuristic for the maximum
  clique problem on gpus and clusters of workstations,'' in \emph{IEEE
  International Conference on Electro-Information Technology , EIT 2013}, 2013,
  pp. 1--6.

\bibitem{daniluk:19:mcgpu}
\BIBentryALTinterwordspacing
P.~Daniluk, G.~Firlik, and B.~Lesyng, ``Implementation of a maximum clique
  search procedure on cuda,'' \emph{Journal of Heuristics}, vol.~25, no.~2, p.
  247–271, Apr. 2019. [Online]. Available:
  \url{https://doi.org/10.1007/s10732-018-9393-x}
\BIBentrySTDinterwordspacing

\end{thebibliography}
